\documentclass[letterpaper,11pt]{article}
\usepackage{jheppub}
\usepackage{epstopdf}
\usepackage{slashed}
\usepackage{graphicx}
\usepackage{amsmath}
\usepackage{multirow}
\usepackage{color}
\usepackage[utf8]{inputenc}
\usepackage[normalem]{ulem}
\usepackage[T1]{fontenc}
\usepackage{cancel}

\usepackage{graphicx}
\usepackage{dcolumn}
\usepackage{bm}

\usepackage{amsmath}
\usepackage{amsfonts}
\usepackage{amssymb}
\usepackage{color}

\newcommand{\beq}{\begin{equation}}
\newcommand{\eeq}{\end{equation}}
\newcommand{\bea}{\begin{eqnarray}}
\newcommand{\eea}{\end{eqnarray}}
\newcommand{\ba}{\begin{array}}
\newcommand{\ea}{\end{array}}

\def\m1{M_1}
\def\m2{M_2}
\def\m3{M_3}

\def\ch10{\tilde \chi^0_1}

\def\to{\rightarrow}

\newcommand{\lsim}{\mathrel{\mathop{\kern 0pt \rlap
  {\raise.2ex\hbox{$<$}}}
  \lower.9ex\hbox{\kern-.190em $\sim$}}}
\newcommand{\gsim}{\mathrel{\mathop{\kern 0pt \rlap
  {\raise.2ex\hbox{$>$}}}
  \lower.9ex\hbox{\kern-.190em $\sim$}}}
\newcommand{\RN}[1]{%
  \textup{\uppercase\expandafter{\romannumeral#1}}%
}

\title{\Large{\bf Axion-Like Particles at High Energy Muon Colliders\\
$-$ A White paper for Snowmass 2021}}

\author[a]{Tao Han,}
\author[b]{Tong Li,}
\author[c]{and Xing Wang}

\affiliation[a]{PITT PACC, Department of Physics and Astronomy, University of Pittsburgh, Pittsburgh, PA 15217, USA}
\affiliation[b]{School of Physics, Nankai University, Tianjin 300071, China}
\affiliation[c]{Department of Physics, University of California at San Diego, La Jolla, CA 92093, USA}

\preprint{PITT-PACC-2202}

\abstract{
We study the discovery potential for heavy axion-like particles (ALPs) and the perspectives for determining their coupling properties at a muon collider. Focusing on their couplings to the Standard Model (SM) gauge bosons $\gamma, Z, W^\pm$, we show that a high-energy muon collider can substantially extend the mass coverage, essentially reaching the kinematic limit of the collider energy. The unique kinematics allow for non-ambiguous determination of the individual coupling strengths. The associated production via $\mu^+\mu^-$ annihilation and the VBF processes with the tagged outgoing muons can be utilized to verify the CP property of the ALPs. We illustrate our results for a muon collider running at 3 TeV and 10 TeV.
}

\makeatletter
\gdef\@fpheader{}
\makeatother

\begin{document}

\maketitle

\noindent
\section{Introduction}

Axion-like particles (ALPs) exist in a variety of theories beyond the Standard Model (SM)~\cite{Kim:1986ax,Kuster:2008zz}.
They are CP-odd scalars, pseudo-Nambu-Goldstone bosons associated with a spontaneous global U(1) symmetry breaking, and singlets under the SM charges.
The best known example is the QCD axion~\cite{Peccei:1977hh,Peccei:1977ur,Weinberg:1977ma,Wilczek:1977pj} (for a recent review see Ref.~\cite{DiLuzio:2020wdo}), originally proposed to solve the strong CP problem~\cite{Baluni:1978rf,Crewther:1979pi,Kim:1979if,Shifman:1979if,Dine:1981rt,Zhitnitsky:1980tq,Baker:2006ts,Pendlebury:2015lrz}. It is soon realized that an ALP may appear in many theoretical constructions, such as composite models~\cite{Kim:1984pt,Choi:1985cb,Kaplan:1985dv,Randall:1992ut,Dobrescu:1996jp,Redi:2016esr,Kobakhidze:2016wmv,Dvali:2016eay,Fukuda:2017ylt,DiLuzio:2017tjx,Lillard:2017cwx,DiLuzio:2017pfr,Gaillard:2018xgk,Lillard:2018fdt,Anastasopoulos:2018uyu,Gavela:2018paw,Lee:2018yak}, extra dimension models~\cite{Dienes:1999gw,Hill:2002kq,Choi:2003wr,Flacke:2006ad,Cox:2019rro,Bonnefoy:2020llz,Yamada:2021uze}, Grand Unification models~\cite{Wise:1981ry,Lazarides:1981kz,Georgi:1981pu,Rubakov:1997vp,Co:2016xti,Lee:2016wiy,Boucenna:2017fna,Daido:2018dmu,DiLuzio:2018gqe,Ernst:2018rod,FileviezPerez:2019fku,FileviezPerez:2019ssf,Bajc:2005zf,Altarelli:2013aqa,Babu:2015bna,Ernst:2018bib,Coriano:2019vjl,DiLuzio:2020qio,Coriano:2017ghp,Chen:2021haa} and superstring theories~\cite{Witten:1984dg,Kallosh:1995hi,Svrcek:2006yi}. Due to the vastly different theoretical motivations and incarnations, the mass ($m_a$) and the scale ($\Lambda$) associated with the ALP physics can be drastically different~\cite{Dimopoulos:1979pp,Tye:1981zy,Dine:1981rt,Holdom:1982ex,Kaplan:1985dv,Srednicki:1985xd,Flynn:1987rs,Kamionkowski:1992mf,Berezhiani:2000gh,Hsu:2004mf,Hook:2014cda,Alonso-Alvarez:2018irt,Hook:2019qoh}, ranging from a light axion in sub-eV~\cite{Kim:1979if,Shifman:1979if,Dine:1981rt,Zhitnitsky:1980tq,Turner:1989vc} to a heavy one in TeV~\cite{Rubakov:1997vp,Fukuda:2015ana,Gherghetta:2016fhp,Dimopoulos:2016lvn,Chiang:2016eav,Gaillard:2018xgk,Gherghetta:2020ofz}, and even beyond~\cite{Kobakhidze:2016rwh,Agrawal:2017ksf,Gherghetta:2020keg}. Therefore, the experimental search for the ALPs would require rather different techniques and facilities. Any experimental observation for such a state, on the other hand, would significantly extend our knowledge for physics beyond the SM~\cite{Kuster:2008zz,DiLuzio:2020wdo,Choi:2020rgn}.

High-energy colliders have been the primary tool for discoveries in the past decades. Once reaching a new energy threshold, the collider experiments will most effectively reveal the mass and interactions of the new states at the energy frontier.
The same approach has been applied to the searches for ALPs at the CERN Large Hadron Collider (LHC)~\cite{Jaeckel:2012yz,Brivio:2017ije,Bauer:2017ris,Mariotti:2017vtv,Bauer:2018uxu,Gavela:2019cmq,Carmona:2021seb,Florez:2021zoo,Wang:2021uyb,Ren:2021prq} and future lepton colliders such as ILC, FCC-ee and CLIC~\cite{Bauer:2018uxu,Liu:2017zdh,Zhang:2021sio,Cacciapaglia:2021aqz,Steinberg:2021wbs} (see Ref.~\cite{dEnterria:2021ljz} and references therein for the current status of ALPs search at LEP, Belle-II, Tevatron and LHC).
High-energy colliders have been the primary tool for discoveries at the energy frontier.
After the discovery of the Higgs boson at the LHC, the luminosity upgrade of the LHC will take the lead in searching for new physics beyond the Standard Model~\cite{Apollinari:2015wtw}. There also have been considerations to construct next generaton of hadron colliders of the order of 100 TeV in c.m.~energy~\cite{FCC:2018byv,CEPC-SPPCStudyGroup:2015csa}.

Recently, high energy muon colliders have gained much attention in the community after the endorsement of its R$\&$D by the European strategy and the subsequent formation of the Muon Collider Collaboration~\cite{muCcoll:2020}.
The recent technological development~\cite{MICE:2019jkl,Delahaye:2019omf,Bartosik:2020xwr} has encouraged the community to consider the high-energy option at multi-TeV, thus provides tremendous opportunities to produce and discover new heavy EW particles. An optimistic scheme is to target on a high integrated luminosity and to scale it with energy quadratically as
\begin{equation}
\label{eq:lumi}
  \mathcal{L}=\left(\frac{\sqrt{s}}{10\ {\rm TeV}}\right)^2 10~\textrm{ab}^{-1}.
\end{equation}
In particular, we consider two benchmark choices of the collider energies and the corresponding integrated luminosities for illustration,
\beq
\sqrt{s} = 3\ {\rm and}\ 10\ {\rm TeV},\quad {\mathcal L} = 1\  {\rm and}\ 10\ {\rm ab}^{-1} .
\label{eq:para}
\eeq
The first choice corresponds to a comparative benchmark associated with the CLIC~\cite{Linssen:2012hp}. The second choice is the high energy option targeted for a future muon collider.

Motivated by this exciting perspective, we perform a first study on the reach of muon colliders for the ALPs.
Based on the general parameterization of the ALP couplings to the SM gauge bosons, we calculate the ALP signal cross sections, their discovery potential, and spin-parity property determination. We demonstrate that a muon collider has the double-advantage for reaching a higher mass threshold via the $\mu^+\mu^-$ direct annihilation, and for offering multiple production channels via vector boson fusion processes. The clean experimental environment and the well-defined kinematics in leptonic collisions provide the great opportunity for determining the ALPs properties.

The rest of this paper is organized as follows. In \autoref{sec:ALPs}, we set up our theoretical framework for the ALPs  interactions with the SM gauge bosons,  present the current search bounds on the theory parameters of the ALP mass and couplings, and comment on the prospects for searches from the other colliders.
In \autoref{sec:ALPPheno}, we present our numerical analyses for the search and test of the ALP properties at a muon collider with the two sets of energy and luminosity benchmarks.
We summarize our results and discuss directions for further exploration in \autoref{sec:Sum}.

\section{General Interactions for ALPs}
\label{sec:ALPs}

Starting from the SM, we introduce a generic massive CP-odd scalar  denoted by $a$, presumably a pseudo Nambu-Goldstone boson $-$ an axion-like particle ALP associated with a global U(1) symmetry spontaneously broken above the electroweak scale.
Besides the kinetic term for ALP, the most general effective Lagrangian for bosonic ALP interactions is composed of four dimension-five operators~\cite{Brivio:2017ije}, respecting the SM gauge symmetry
\begin{eqnarray}
\mathcal{L}_{eff}
= C_{\tilde{G}}\mathcal{O}_{\tilde{G}} + C_{\tilde{B}}\mathcal{O}_{\tilde{B}} + C_{\tilde{W}}\mathcal{O}_{\tilde{W}} +C_{a\Phi}\mathcal{O}_{a\Phi}\;,
\label{eq:effLagrangian}
\end{eqnarray}
with
\begin{eqnarray}
&&\mathcal{O}_{\tilde{G}}\equiv -{a\over f_a}G^i_{\mu\nu}\tilde{G}_i^{\mu\nu}\;,~~\mathcal{O}_{\tilde{W}}\equiv -{a\over f_a}W^j_{\mu\nu}\tilde{W}_j^{\mu\nu}\;,\\
&&\mathcal{O}_{\tilde{B}}\equiv -{a\over f_a}B_{\mu\nu}\tilde{B}^{\mu\nu}\;,~~\mathcal{O}_{a\Phi}\equiv i{\partial^\mu a\over f_a}(\Phi^\dagger \overleftrightarrow{D}_\mu \Phi )\;,
\end{eqnarray}
where $\Phi$ is the SM Higgs doublet, $G_{\mu\nu}^i\ (i=1,\cdots,8)$, $W_{\mu\nu}^j\ (j=1,2,3)$ and $B_{\mu\nu}$ denote the field strength tensors of $SU(3)_c$, $SU(2)_L$ and $U(1)_Y$ gauge fields, respectively, and the dual field strengths are defined as $\tilde{X}^{\mu\nu}\equiv {1\over 2}\epsilon^{\mu\nu\alpha\beta}X_{\alpha\beta}$ with $\epsilon^{0123}=1$. The new physics scale is denoted by the constant $f_a$. After the electroweak symmetry breaking, except the Yukawa-axion coupling induced by $\mathcal{O}_{a\Phi}$, the interactions between ALP and the physical SM gauge bosons are
\begin{eqnarray}
\mathcal{L}_{eff}
\supset
&-&{g_{agg}\over 4} a G^a_{\mu\nu}\tilde{G}_a^{\mu\nu} -{g_{a\gamma\gamma}\over 4} a F_{\mu\nu}\tilde{F}^{\mu\nu}-{g_{a\gamma Z}\over 4} a F_{\mu\nu}\tilde{Z}^{\mu\nu}
\nonumber \\
&-&{g_{aZZ}\over 4} a Z_{\mu\nu}\tilde{Z}^{\mu\nu}-{g_{aWW}\over 4} a W_{\mu\nu}\tilde{W}^{\mu\nu}\;,
\label{eq:Lagrangian}
\end{eqnarray}
where
\begin{eqnarray}
&&g_{agg}={4\over f_a}C_{\tilde{G}}\;,~~g_{a\gamma\gamma}={4\over f_a}(s_\theta^2 C_{\tilde{W}}+c_\theta^2 C_{\tilde{B}})\;,~~g_{aZZ}={4\over f_a}(c_\theta^2 C_{\tilde{W}}+s_\theta^2 C_{\tilde{B}})\;,\nonumber\\
&&g_{a\gamma Z}={8\over f_a}s_\theta c_\theta (C_{\tilde{W}}-C_{\tilde{B}})\;,~~g_{aWW}={4\over f_a}C_{\tilde{W}}\;,
\label{eq:couplings}
\end{eqnarray}
with $s_\theta$ ($c_\theta$) being the sine (cosine) of the weak mixing angle $\theta_W$. Below we assume the absence of the gluonic contribution and the Yukawa-axion coupling $C_{\tilde{G}}=C_{a\Phi}=0$ to focus on the EW sector only for simplicity. The Feynman rule between the ALP and two SM gauge bosons ($V_1$ and $V_2$) turns out to be\footnote{The complete Feynman rules from the bosonic ALP effective Lagrangian can be found in the Appendix B of Ref.~\cite{Brivio:2017ije}}
\begin{eqnarray}
-i g_{a V_1 V_2}~p_{V_1 \alpha} p_{V_2 \beta}~\epsilon^{\mu\nu\alpha\beta} \;,
\label{eq:tensor}
\end{eqnarray}
with the momenta ($p_{V_1}$, $p_{V_2}$) flowing inwards in the vertices.
The independent model parameters to be studied are $C_{\tilde{W}}/f_a$ and $C_{\tilde{B}}/f_a$ in this
formalism.

\section{ALP phenomenology at Muon Colliders}
\label{sec:ALPPheno}

The future high energy Muon colliders have great potential in probing ALP at multi-TeV scales. In the section, we study the ALP phenomenology at high energy muon colliders.
In particular, we consider three benchmark scenarios where
\begin{eqnarray}
&(\RN{1})~~~&C_{\tilde{W}}=C_{\tilde{B}}\neq 0\; ;\\
&(\RN{2})~~~&C_{\tilde{W}} = 0,\quad C_{\tilde{B}}\neq 0\; ;\\
&(\RN{3})~~~&C_{\tilde{W}}\neq 0,\quad C_{\tilde{B}} = 0\;.
\end{eqnarray}

\subsection{Production}
The ALP can be produced at muon colliders through two different topologies,  the associated production via $\mu^+\mu^-$ annihilation and the electroweak vector-boson-fusion (VBF). In this section, we study the signature of different production modes and their projected sensitivities at muon colliders.

\subsubsection{Associated production}
\label{sec:anni}

\begin{figure}
    \centering
    \includegraphics[width=0.32\textwidth]{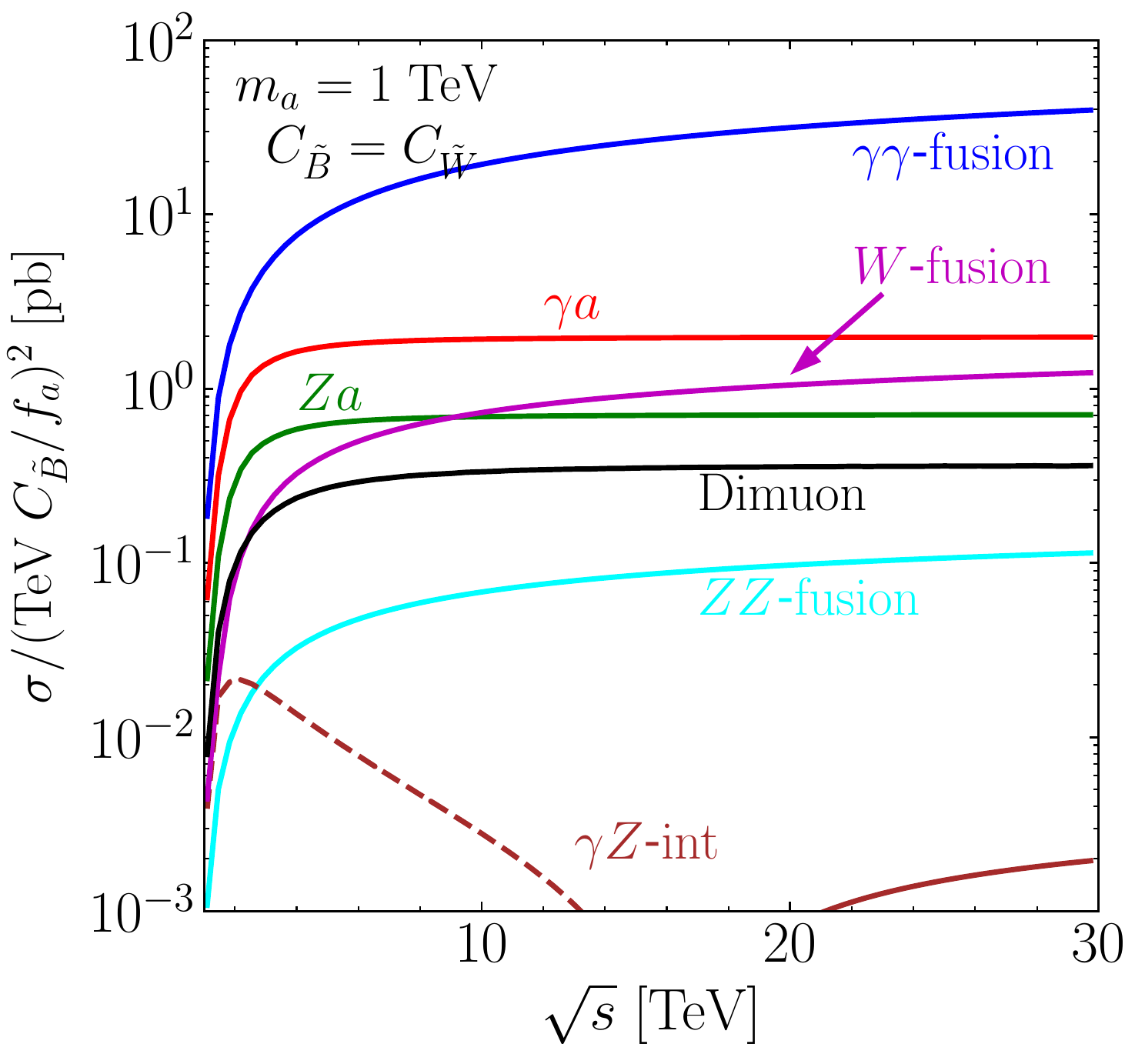}
    \includegraphics[width=0.32\textwidth]{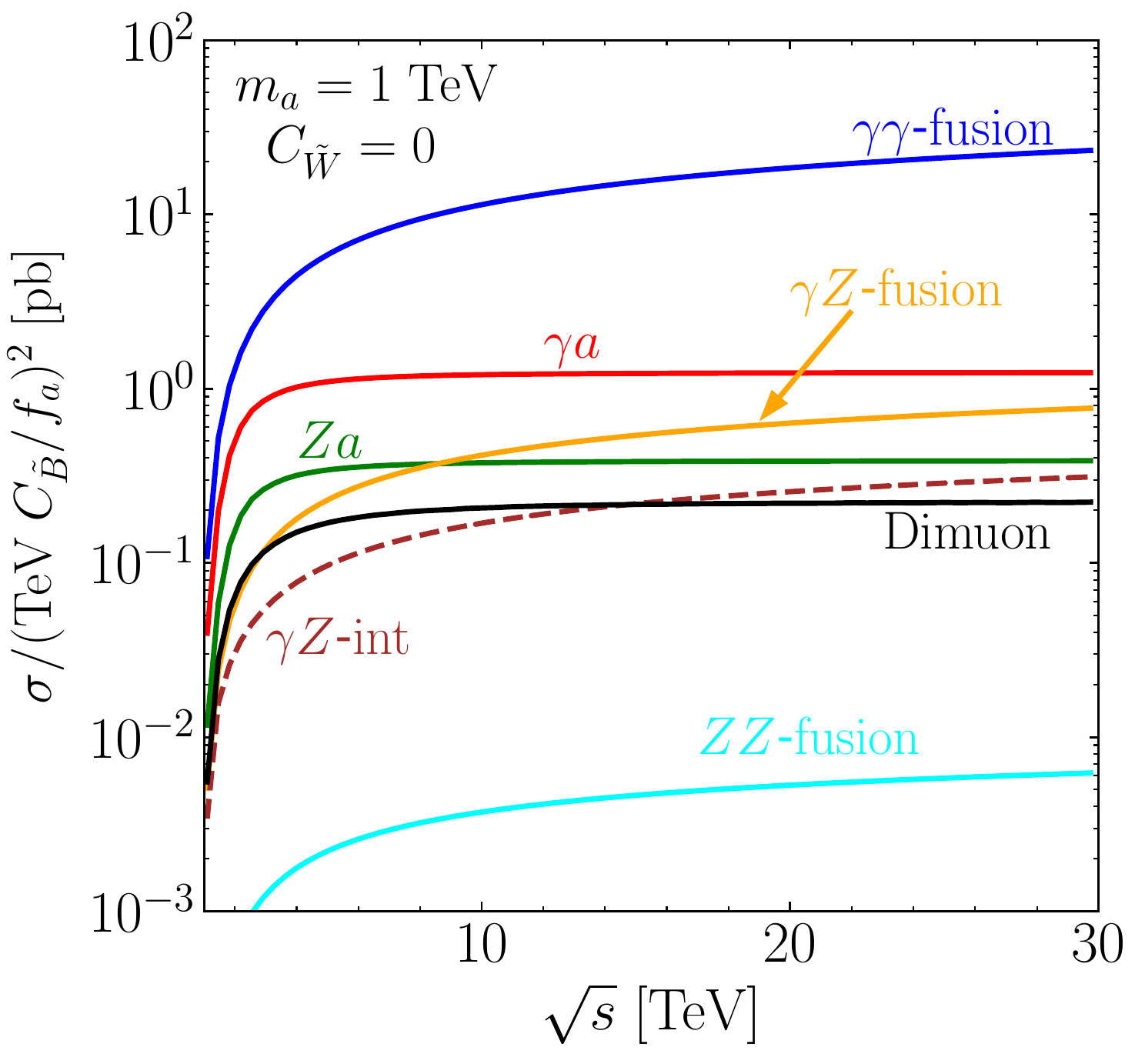}
    \includegraphics[width=0.32\textwidth]{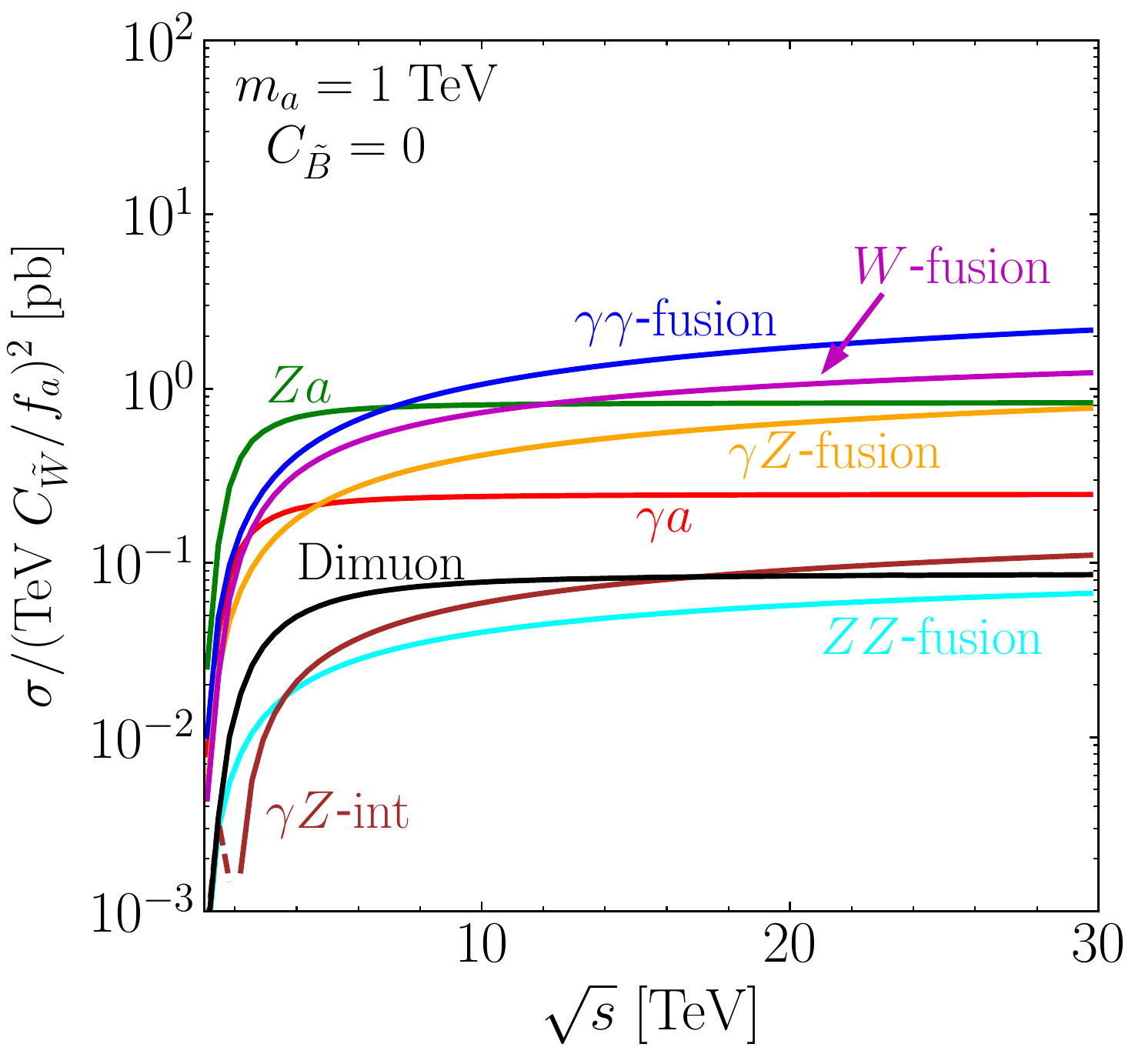}
    \caption{The cross sections of different production modes from the direct annihilation $\mu^+\mu^-\to \gamma a, Z a$, and the VBF processes, as a function of $\sqrt{s}$ for $m_a = 1$ TeV.
    The dashed brown curves indicate the $\gamma Z$ destructive interference.}
    \label{fig:xsec_s}
\end{figure}

We first consider the ALP production associated with an electroweak vector boson through $\mu^+\mu^-$ annihilation processes
\begin{eqnarray}
\mu^+\mu^-\to V a,\quad V=\gamma, Z\;.
\end{eqnarray}
Each production above is in general determined by the interference of two diagrams induced by two ALP couplings which are $g_{a\gamma\gamma}, g_{a\gamma Z}$ (for $\gamma a$ production) and $g_{a\gamma Z}, g_{aZZ}$ (for $Z a$ production).
Figures \ref{fig:xsec_s} and \ref{fig:xsec_m} show the production cross sections as a function of $\sqrt{s}$ and $m_a$, respectively, where the associated productions are represented by red ($\gamma a$) and green ($Za$) lines. The cross sections are normalized by $C_{\tilde{W},\tilde{B}}^2/f_a^2$. For the case with $C_{\tilde{W}}=C_{\tilde{B}}$, in particular, the coupling $g_{a\gamma Z}$ vanishes and there appears only one relevant diagram.
The momenta-dependence of the dimension-5 operators is cancelled by the $s$-channel propagator, and this cancellation leads to a constant behavior of cross section at high energies as shown by the red ($\gamma a$) and green ($Z a$) lines in Fig.~\ref{fig:xsec_s}
\begin{eqnarray}
\sigma_{V_1^\ast V_2 a}\propto g_{aV_1V_2}^2\sim C_{\tilde{W},\tilde{B}}^2/f_a^2 \;,
\end{eqnarray}
where $V_1^\ast$ denotes the gauge boson propagator in the $s$-channel and $V_2$ is the one in final states. The falling behavior as a function of $m_a$ in the panels of Fig.~\ref{fig:xsec_m} only comes from the suppression of phase space.

\subsubsection{Vector boson fusion production}
\label{sec:fusion}

\begin{figure}
    \centering
    \includegraphics[width=0.32\textwidth]{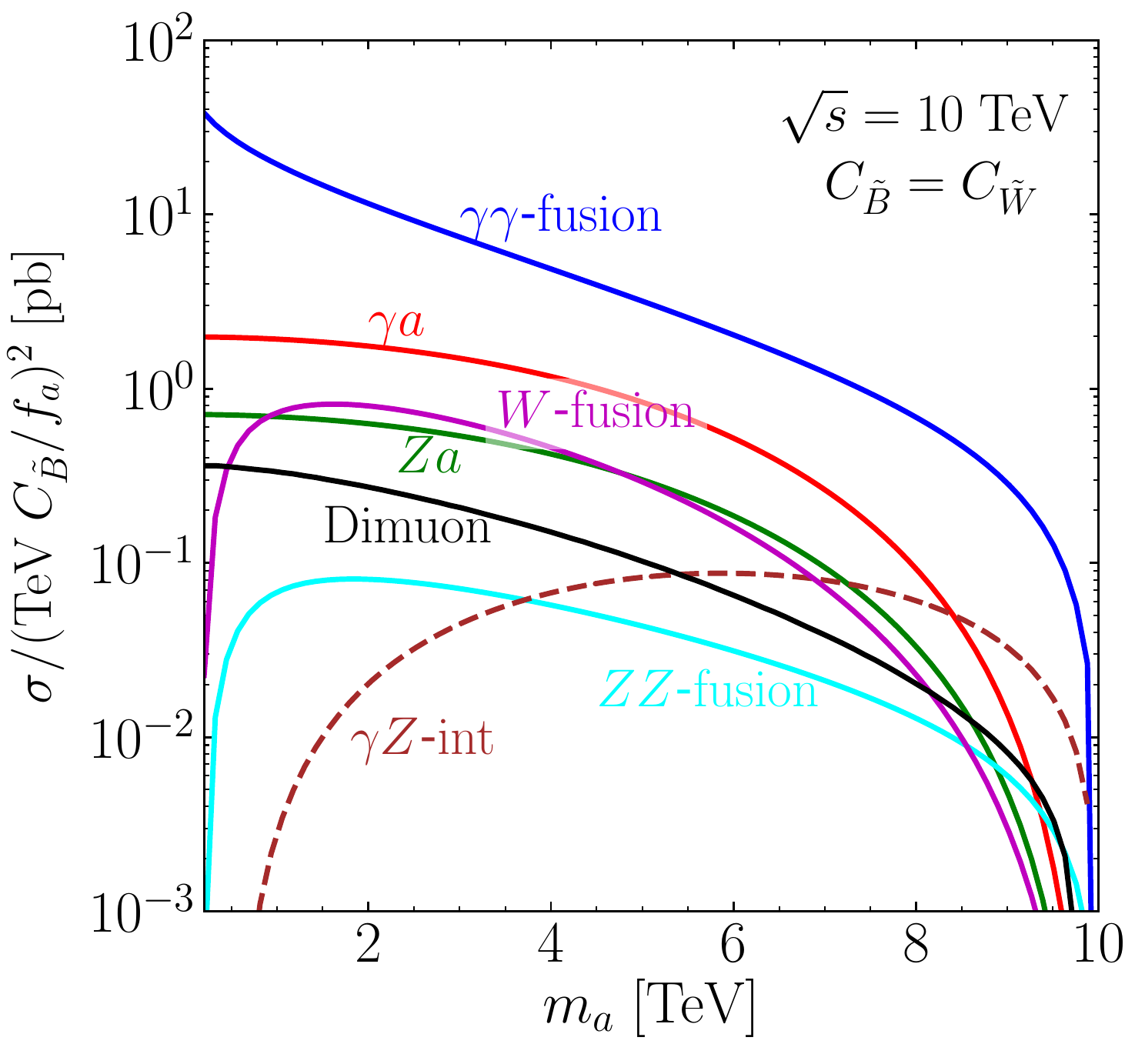}
    \includegraphics[width=0.32\textwidth]{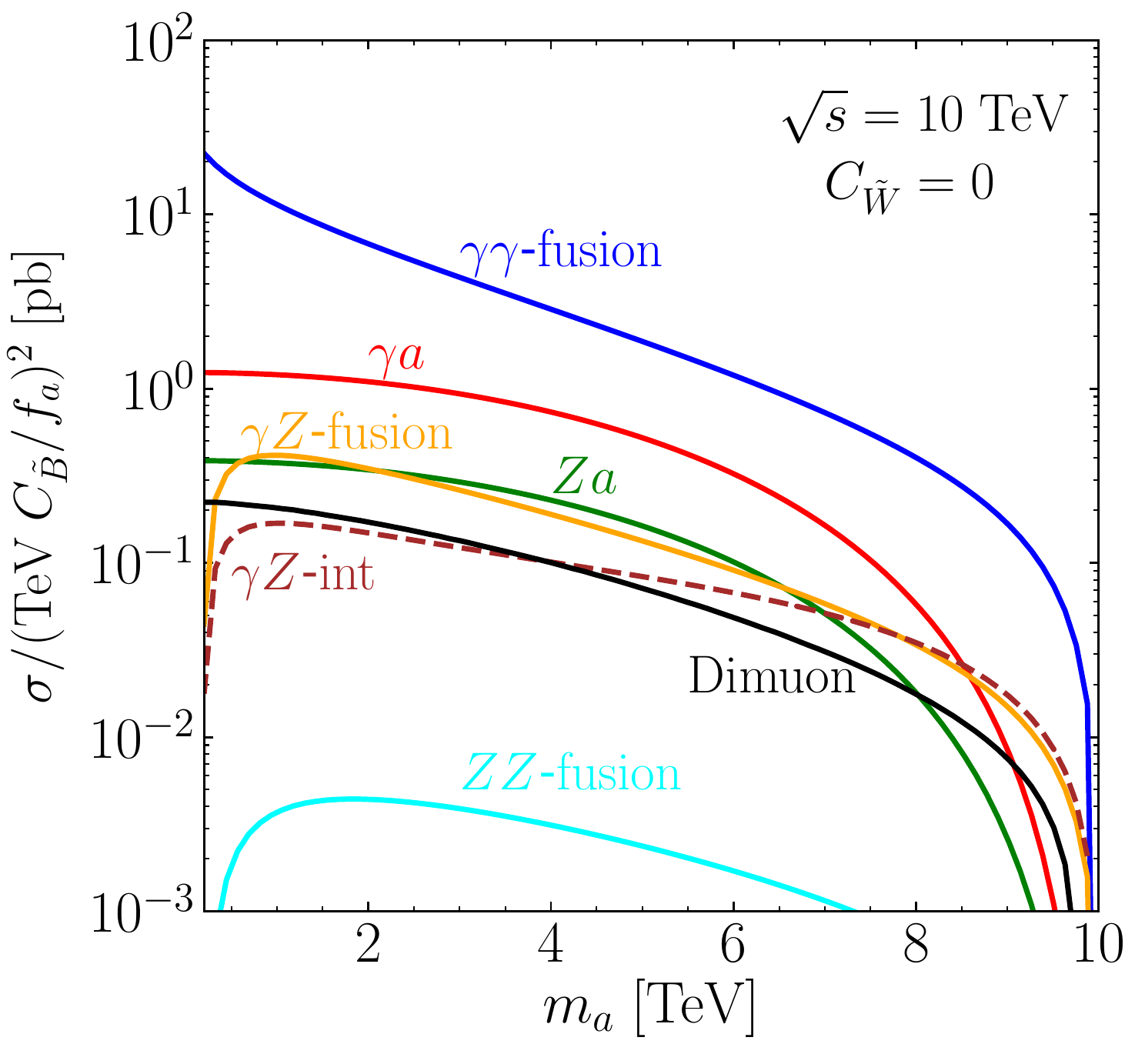}
    \includegraphics[width=0.32\textwidth]{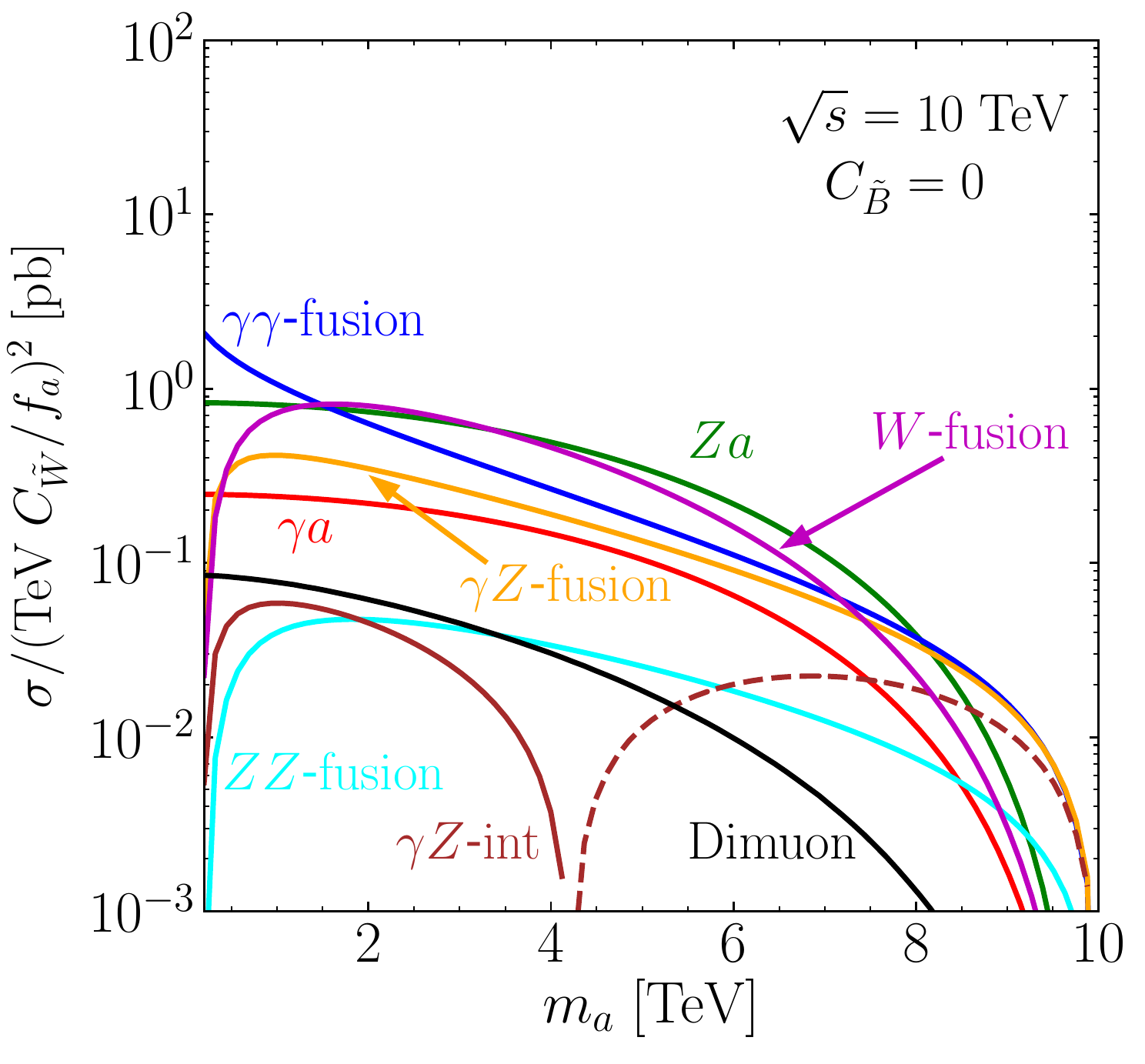}
    \caption{The cross sections of different production modes from the direct annihilation $\mu^+\mu^-\to \gamma a, Z a$, and the VBF processes, as a function of $m_a$ at $\sqrt{s} = 10$ TeV muon colliders.
    The dashed brown curves indicate the $\gamma Z$ destructive interference.}
    \label{fig:xsec_m}
\end{figure}

At higher colliding energies, the productions of ALPs through VBF processes
\begin{equation}
\gamma\gamma,\ ZZ,\ \gamma Z,\ WW\to a
\end{equation}
become increasingly important. For c.m.~energies $\sqrt{s} \gg m_W$, the initial muon beams substantially radiate EW gauge bosons under an approximately unbroken SM gauge symmetry. This becames particularly problematic for photon-photon fusions, where the collinear singularity is regulated by the tiny muon mass and and leads to large logarithm ${\rm ln}(s/m_{\mu}^2)$. We treat the vector bosons as initial state partons, and calculate the VBF cross sections, by utilizing the leading-order framework of electroweak parton distribution functions (EW PDFs)~\cite{Han:2020uid} with a dynamical scale $Q=\sqrt{\hat{s}}/2$, where $\sqrt{\hat{s}}$ is the partonic c.m.~energy.
For the initial gauge boson partons $V_i$ and $V_j$, the VBF production cross section can be factorized as the product of the parton luminosity $d\mathcal{L}_{ij}/d\tau$ and the partonic cross section $\hat{\sigma}$
\begin{eqnarray}
\sigma(\ell^+\ell^-\to F+X)&=&\int^1_{\tau_0}d\tau \sum_{ij}{d\mathcal{L}_{ij}\over d\tau} \hat{\sigma}(V_i V_j\to F)\;,\nonumber \\
{d\mathcal{L}_{ij}\over d\tau}&=&{1\over 1+\delta_{ij}}\int^1_\tau {d\xi\over \xi}\Big[f_i(\xi,Q^2)f_j({\tau\over \xi},Q^2)+(i\leftrightarrow j)\Big]\;,
\end{eqnarray}
where $F (X)$ denotes an exclusive final state (the underlying remnants), $f_i(\xi,Q^2)$ is the EW PDF for vector $V_i$ with $Q$ being the factorization scale, $\tau_0=m_F^2/s$ and $\tau=\hat{s}/s$. By summing over all gauge boson initial states, one can obtain the total cross section of the ``inclusive'' production processes.
The cross section is enhanced by collinear logarithm ${\rm ln}(\hat{s}/m_\mu^2)$ for photon or ${\rm ln}(\hat{s}/m_V^2)$ for massive gauge boson $V$ at high beam energies. In contrast to the constant behavior in the associated production, such logarithmic enhancement can be seen in Fig.~\ref{fig:xsec_s} where we choose the factorization scale as $Q=\sqrt{\hat{s}}/2$. Moreover, suppose the momenta of incoming two gauge bosons ($p_{V_1 \alpha}$, $p_{V_2 \beta}$) are longitudinally back-to-back along the beam direction, the totally antisymmetric tensor $\epsilon^{\mu\nu\alpha\beta}$ in the ALP vertex determines the polarization of them ($\varepsilon_{V_1 \mu}$, $\varepsilon_{V_2 \nu}$) to be transverse. Thus, each VBF production is dominated by the transverse gauge boson ($\gamma, W_T, Z_T$) fusion.

In Figs.~\ref{fig:xsec_s} and \ref{fig:xsec_m}, we show the production cross sections of each individual fusion channels:
\begin{eqnarray}
&\gamma \gamma \rightarrow a~~&{\rm (blue)}\; ,\nonumber\\
&Z Z \rightarrow a~~&{\rm (cyan)}\; ,\nonumber\\
&\gamma Z \rightarrow a~~&{\rm (orange)}\; ,\nonumber\\
&W^+ W^- \rightarrow a~~&{\rm (magenta)}\; .\nonumber
\end{eqnarray}
Note that the $\gamma$- and $Z$-initiated fusion processes are physically indistinguishable, and in principle should be added coherently. For the sake of illustration, we compute such interference, also in the leading-log approximation, which is shown using brown lines in Figs.~\ref{fig:xsec_s} and \ref{fig:xsec_m}. The solid and dashed lines correspond to constructive and destructive interference, respectively. While the interference effects are usually orders of magnitude smaller, they do become important for heavy axion masses. As shown in Figs.~\ref{fig:xsec_s} and \ref{fig:xsec_m}, when the $g_{a\gamma\gamma}$ coupling is not suppressed, the $\gamma\gamma$-fusion always dominates, due to its ${\rm ln}(s/m_{\mu}^2)$ enhancement, as in scenarios $(\RN{1})$ and $(\RN{2})$. However, as in scenario $(\RN{3})$, the $g_{a\gamma\gamma}$ coupling is suppressed by $s_\theta^2$ and the $\gamma\gamma$-fusion become significantly smaller.

In comparison to the inclusive VBF production, for the $\gamma$- and $Z$-initiated fusion processes, one can also consider the exclusive VBF processes by requiring the outgoing $\mu^+\mu^-$ to be observable in the detector coverage
\begin{equation}
    10^\circ < \theta_{\mu^\pm} < 170^\circ.
    \label{eq:dimuon_1}
\end{equation}
Such requirement would greatly suppress the production cross section, since the outgoing muons tend to be collinear to the beamline, especially for the $\gamma\gamma$-fusion. In addition, we also require
\begin{equation}
    m_{\mu^+\mu^-} > 200~{\rm GeV},
    \label{eq:dimuon_2}
\end{equation}
for the exclusive channel to enhance the its VBF topology. As shown by black lines in Figs.~\ref{fig:xsec_s} and \ref{fig:xsec_m} labelled by ``Dimuon'', the cross section for exclusive processes are typically two orders of magnitude smaller than the ones for inclusive processes. Such difference becomes smaller in the case of $C_{\tilde{B}} = 0$, where the $\gamma\gamma$-fusion is suppressed by the relatively small coupling $g_{a\gamma\gamma}$. In spite of smaller cross sections, tagging the outgoing muons can still provide an extra handle for the signal event selection and help to reveal the CP property of the ALPs, as discussed in the next section.

\subsection{Projected bounds}

To estimate the sensitivity, we consider the simplest decay channel as the signal
\begin{equation}
    a \rightarrow \gamma\gamma\; .
\end{equation}
The leading background for the associated production is
\begin{equation}
    \mu^+\mu^-\rightarrow V\gamma\gamma,\quad V=\gamma, Z\; .
    \label{eq:bkg_1}
\end{equation}
For VBF processes, we consider two different signal categories:
\begin{eqnarray}
    &{\rm inclusive:~~}&\mu^+\mu^- \rightarrow a +X\; ,\\
    &{\rm dimuon:~~}&\mu^+\mu^- \rightarrow a + \mu^+\mu^-\; ,
\end{eqnarray}
where the exclusive dimuon channel is as described in the previous subsection, with the cuts in Eqs.~(\ref{eq:dimuon_1}) and (\ref{eq:dimuon_2}) applied. The dominant background for the inclusive production is
\begin{equation}
    \mu^+\mu^-\rightarrow\gamma\gamma\; ,
    \label{eq:bkg_2}
\end{equation}
where the invariant mass of the photon pair is smeared due to the initial-state-radiation (ISR) effect. We simulated such background using WHIZARD~\cite{Kilian:2007gr}. For the exclusive production, the dominant background becomes
\begin{equation}
\mu^+\mu^-\rightarrow\mu^+\mu^-\gamma\gamma\; .
\label{eq:bkg_3}
\end{equation}
We simulate the backgrounds in Eqs.~(\ref{eq:bkg_1}), (\ref{eq:bkg_2}) and (\ref{eq:bkg_3}) using MadGraph~\cite{Alwall:2014hca} at the parton level.

For photon reconstruction, we impose the following basic cuts of transverse momentum, rapidity, and separation on the photons in final states
\begin{eqnarray}
p_{T}(\gamma)> 10~{\rm GeV},~~~|\eta(\gamma)|<2.5,~~~\Delta R_{\gamma\gamma}>0.4\;.
\label{eq:cut1}
\end{eqnarray}
To suppress the continuum background, we further impose the invariant mass on the diphoton resonance
\begin{equation}
    \frac{|m_{\gamma\gamma} - m_a|}{m_a} < 0.05\;.
    \label{eq:cut2}
\end{equation}
We then estimate the significance and evaluate the projected sensitivities.
The local significance is quantified as
\begin{equation}
    N_{\rm SD} = \frac{S}{\sqrt{S+B}}\;,
\end{equation}
where $S$ and $B$ are the numbers of events for the signal and background, respectively. 

\begin{figure}[tb]
\centering
\includegraphics[width=0.32\textwidth]{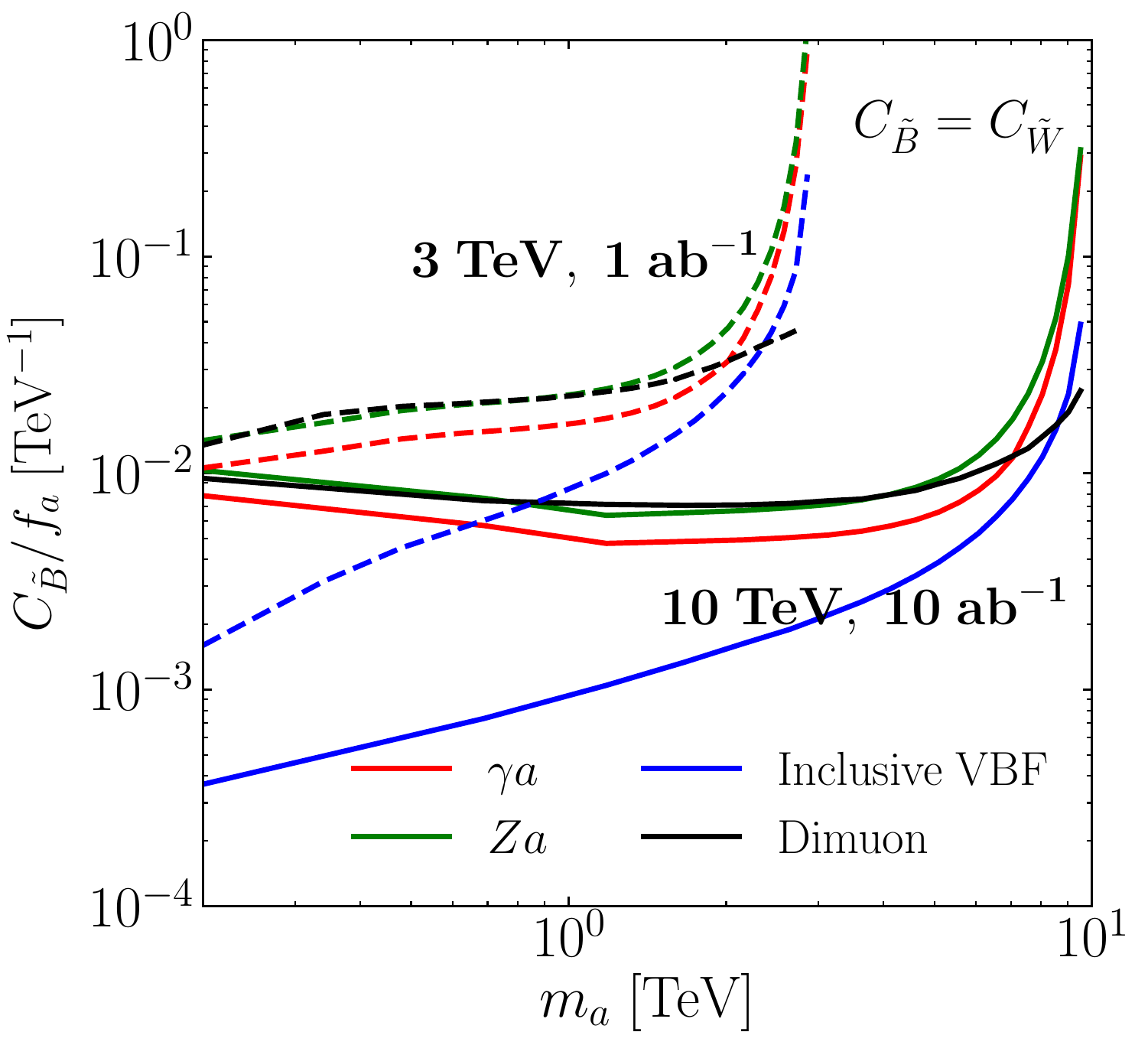}
\includegraphics[width=0.32\textwidth]{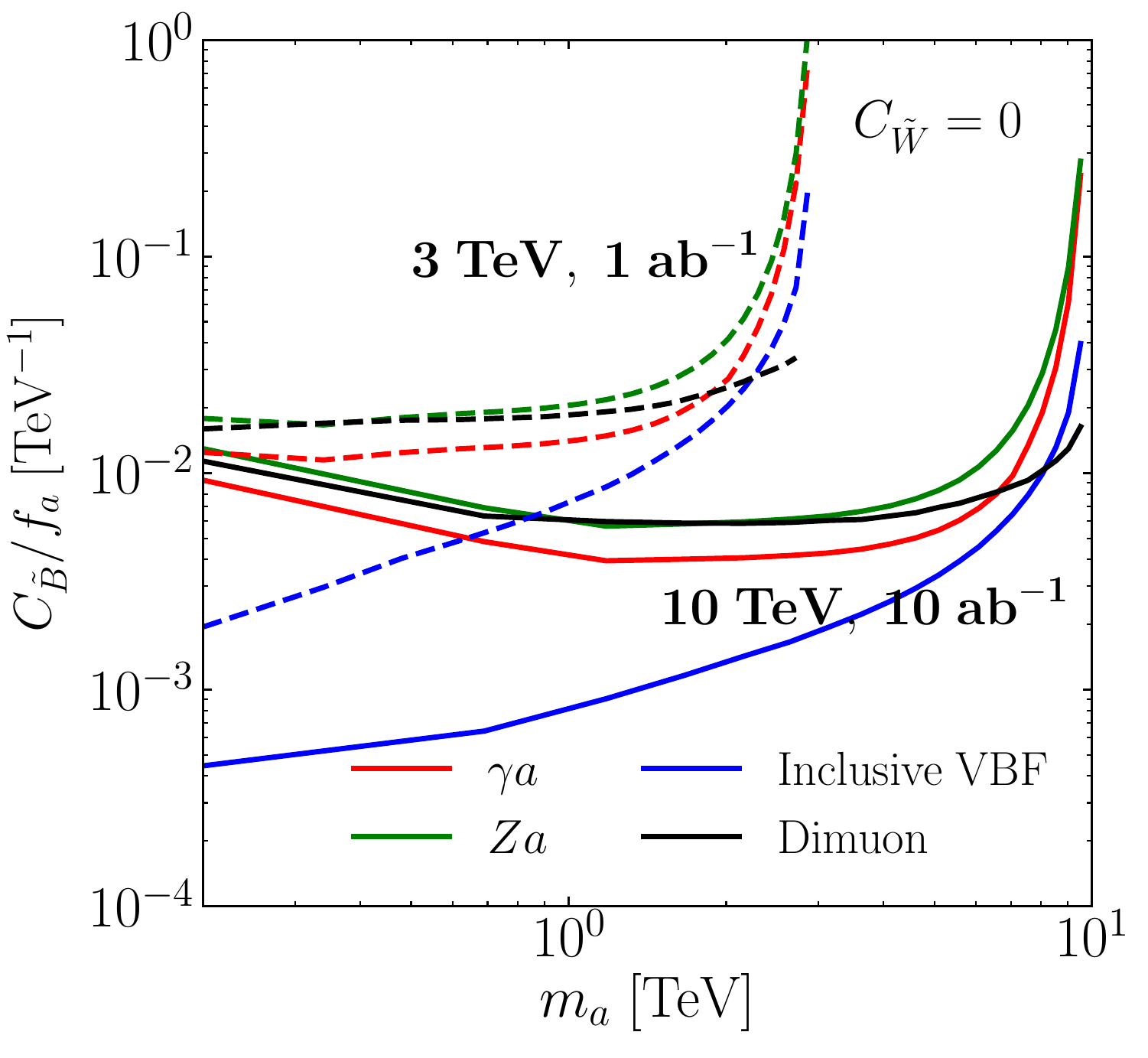}
\includegraphics[width=0.32\textwidth]{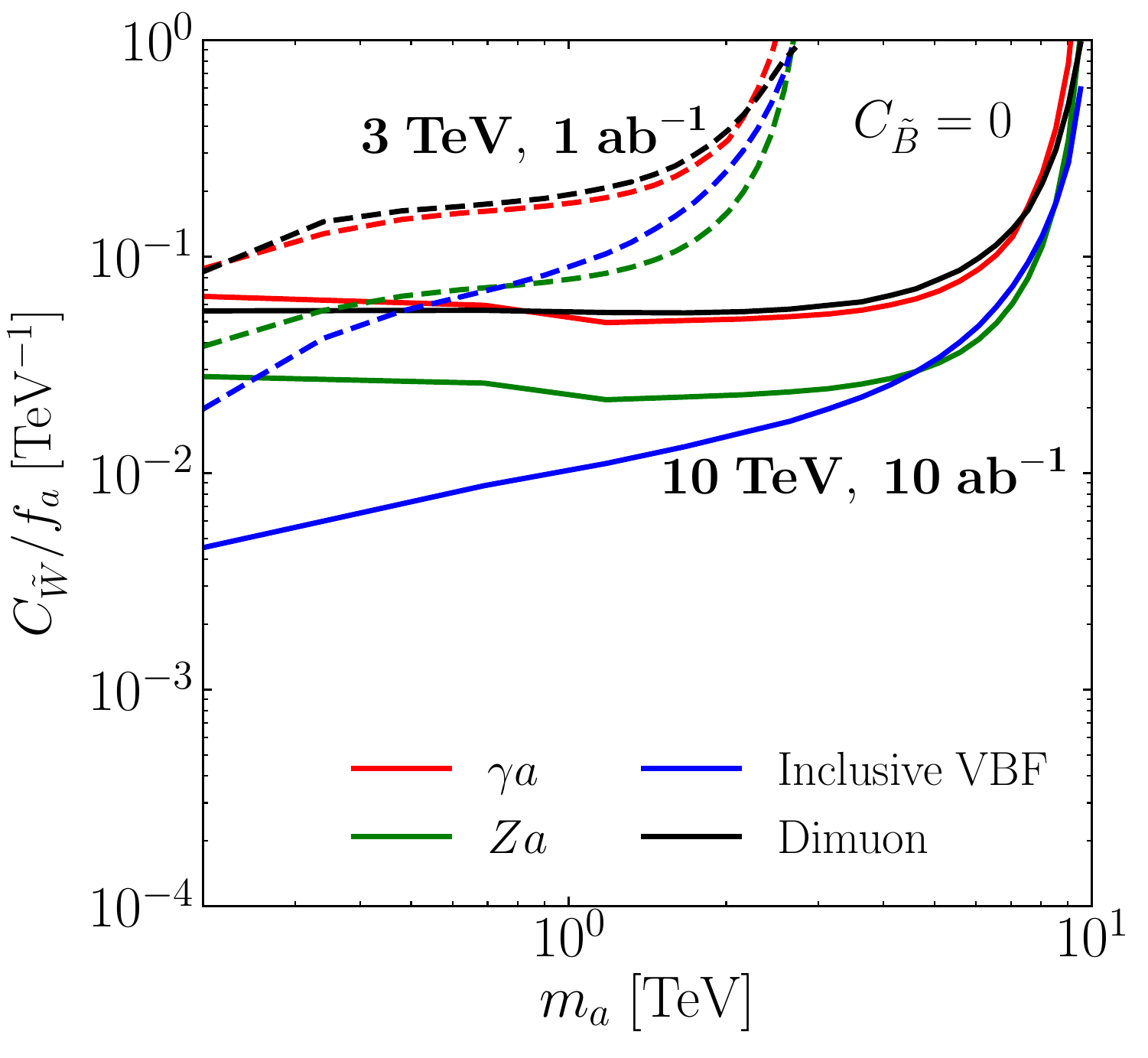}
\caption{Projected $5\sigma$ sensitivity of $C_{\tilde{B}}/f_a$ versus $C_{\tilde{W}}/f_a$. Selection cuts include Eq.~(\ref{eq:cut1}), $|m_{\gamma\gamma} - m_a|/m_a < 5\%$, and $m_{\mu^+\mu^-} > 200$ GeV. The leading backgrounds considered for $\gamma a$, $Za$, inclusive VBF, and Dimuon channels are $\mu^+\mu^-\rightarrow \gamma\gamma\gamma$, $\mu^+\mu^-\rightarrow Z\gamma\gamma$, $\mu^+\mu^-\rightarrow \gamma\gamma$ with ISR, and $\mu^+\mu^-\rightarrow \mu^+\mu^-\gamma\gamma$, respectively.
}
\label{fig:bound}
\end{figure}

In Fig.~\ref{fig:bound}, we show the projected $N_{\rm SD} = 5$ discovery limit on the couplings $C_i/f_a$ as functions of the ALP mass $m_a$, for the two different muon collider benchmarks in Eq.~(\ref{eq:para}).
in dashed and solid lines, respectively.

In scenario $(\RN{1})$ $C_{\tilde{W}}=C_{\tilde{B}}$ and $(\RN{2})$ $C_{\tilde{W}}=0$, the inclusive VBF offers the strongest bounds, especially for relatively light ALP masses. This becomes even more true at $\sqrt{s} = 10$ TeV, due to the large logarithmic enhancement from the photon PDF. The bounds from associated production remain mostly constant for a large range of ALP mass. They become slightly worse at lower masses and higher energy where diphoton from ALP decay turns out to be collimated and boosted techniques are required to reconstruct the resonance. For ALP with mass $m_a = 1$ TeV, the coefficients can be probed as low as $|C_{\tilde{B}}|/f_a \sim 10^{-2}~(10^{-3})$~${\rm TeV}^{-1}$ at $\sqrt{s} = 3~(10)$~TeV.

In scenario $(\RN{3})$ $C_{\tilde{B}}=0$, the suppressed coupling $g_{a\gamma\gamma}$ affects both the $\gamma\gamma$-fusion production and the diphoton branching fraction, resulting in looser bounds. It is worth noting that, at $\sqrt{s} = 3$ TeV, the associated production of $Za$ exceed the VBF processes for $m_a \gtrsim 700$ GeV.

\subsection{Characterizing the CP Property of ALPs at Muon Collider}
\label{sec:Prop}

\begin{figure}[tb]
\centering
\includegraphics[width=0.45\textwidth]{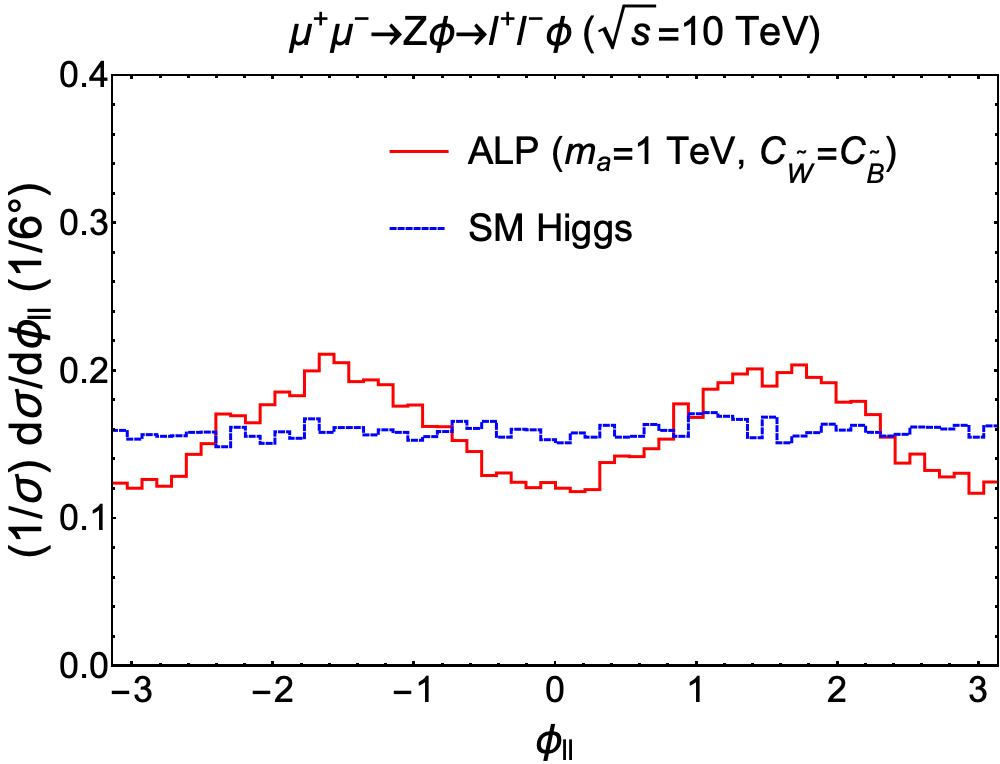}
\includegraphics[width=0.45\textwidth]{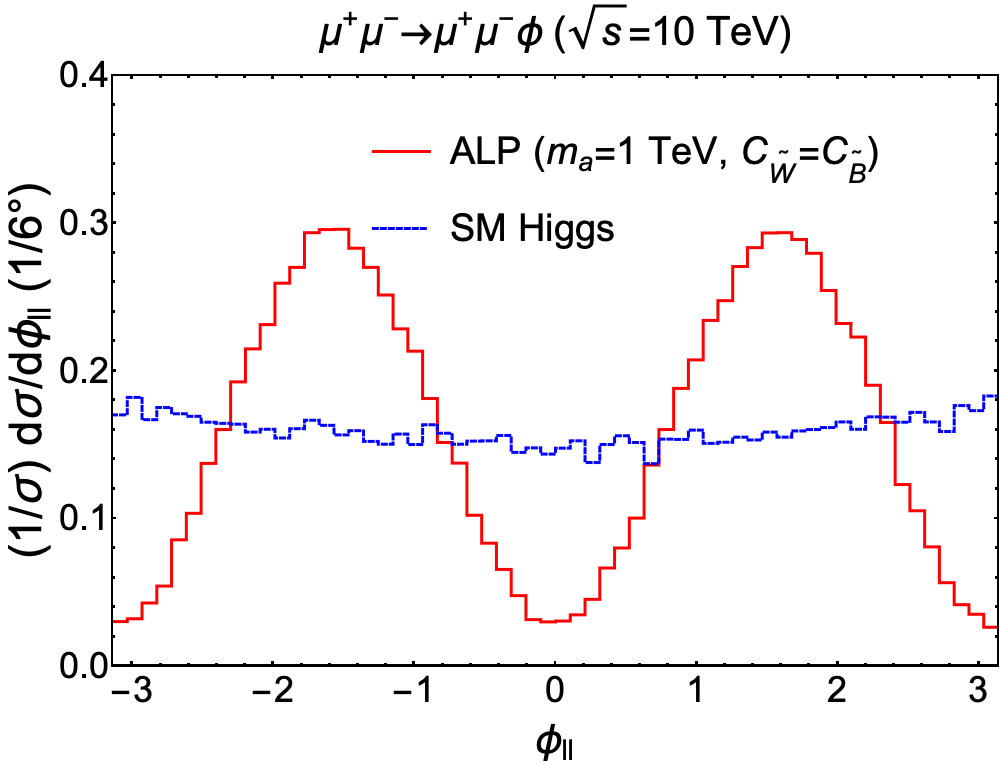}
\caption{The normalized distributions of the observable $\phi_{\ell\ell}$ in $\mu^+\mu^-\to Z \phi\to \ell^+\ell^-\phi$ (left) and $\mu^+\mu^-\to \mu^+ \mu^- \phi$ (right) for $\phi$ as ALP (red) or SM Higgs (blue). We assume $m_a=1$ TeV and $C_{\tilde{W}}=C_{\tilde{B}}\neq 0$ for $\phi$ as the ALP. }
\label{fig:CPDY}
\end{figure}

To verify the CP property of the scalar particle produced at a  muon collider, we first consider the associated production
\begin{eqnarray}
\mu^+\mu^-\to Z\phi\;,
\end{eqnarray}
followed by $Z\to \ell^+\ell^-$ and $\phi=a$ as $0^-$ pseudoscalar ALP or $\phi=h$ as $0^+$ SM Higgs scalar. The azimuthal angle  $\phi_{\ell\ell}$ is defined as the opening angle between the $Z$ production and decay planes. The differential cross section of $\phi_{\ell\ell}$ for ALP becomes
\begin{eqnarray}
\frac{1}{\sigma}\frac{d\sigma}{d\phi_{\ell\ell}} &=& \frac{1}{2\pi}\left(1 - \frac{1}{4}\cos2\phi_{\ell\ell} \right)\;.
\end{eqnarray}
We also examine the following exclusive VBF channels to probe the CP property
\begin{eqnarray}
&&\mu^+\mu^-\to \mu^+ \mu^- \phi\;.
\end{eqnarray}
We apply $m_{\mu\mu}>100$ GeV for final states to enhance the VBF topology and require the following basic cuts
\begin{eqnarray}
p_{T}(\mu)> 10~{\rm GeV},~~~10^\circ<\theta_\mu<170^\circ,~~~\Delta R_{\mu\mu}>0.4\;.
\end{eqnarray}
In this way we require to tag two forward muons, and define $\phi_{\ell\ell}$ as the azimuthal angle between the two planes of the final state muons formed with respect to the beam direction. As seen in Fig.~\ref{fig:CPDY}, the CP-even scalar yields a flat distribution as expected; and the CP-odd scalar exhibits explicitly different angular distribution as governed by the tensor form interaction in Eq.~(\ref{eq:tensor}).

\section{Executive Summary}
\label{sec:Sum}

In this document, we studied the search potential for a heavy ALPs at future high-energy and high-luminosity muon colliders.
We considered the ALP production associated with a neutral electroweak gauge boson ($\gamma,Z$) and the various VBF processes together with the ALP decay into diphoton.

Our main results are summarized as follows.
\begin{itemize}
\item The $\gamma\gamma$ fushion process is dominant in the ALP productions as long as the $U(1)_Y$ gauge coupling is not suppressed.
\item The ALPs can be probed as heavy as the colliding energy threshold above TeV in both associated production and VBF processes.
\item The ALP gauge couplings can be reached as small as $|C_{\tilde{W},\tilde{B}}|/f_a \sim 10^{-2}~(10^{-3})$~${\rm TeV}^{-1}$ at $\sqrt{s} = 3~(10)$~TeV for $m_a=1$ TeV.
\item The associated productions and the VBF processes with the tagged outgoing muons can be utilized to reveal the CP property of the ALPs.
\end{itemize}

We conclude that a muon collider running at high energies with a high luminosity would have great potential in searching for the ALPs, essentially reaching the kinematic limit, and studying their interaction properties owing to the rather clean experimental environment at th lepton colliders.

\begin{acknowledgments}
The work of TH was supported in part by the U.S.~Department of Energy under grant
No.~DE-SC0007914 and in part by the PITT PACC. TL is supported by the National Natural Science Foundation of China (Grant No. 11975129, 12035008) and ``the Fundamental Research Funds for the Central Universities'', Nankai University (Grant No. 63196013). XW was supported by the National Science Foundation under Grant No.~PHY-1915147.
\end{acknowledgments}


\bibliographystyle{JHEP}
\bibliography{refs}

\providecommand{\href}[2]{#2}\begingroup\raggedright\begin{thebibliography}{100}

\bibitem{Kim:1986ax}
J.E.~Kim, \emph{{Light Pseudoscalars, Particle Physics and Cosmology}},
  \href{https://doi.org/10.1016/0370-1573(87)90017-2}{\emph{Phys. Rept.}
  {\bfseries 150} (1987) 1}.

\bibitem{Kuster:2008zz}
M.~Kuster, G.~Raffelt and B.~Beltran, eds., \emph{{Axions: Theory, cosmology,
  and experimental searches. Proceedings, 1st Joint ILIAS-CERN-CAST axion
  training, Geneva, Switzerland, November 30-December 2, 2005}}, vol.~741,
  2008.

\bibitem{Peccei:1977hh}
R.D.~Peccei and H.R.~Quinn, \emph{{CP Conservation in the Presence of
  Instantons}}, \href{https://doi.org/10.1103/PhysRevLett.38.1440}{\emph{Phys.
  Rev. Lett.} {\bfseries 38} (1977) 1440}.

\bibitem{Peccei:1977ur}
R.D.~Peccei and H.R.~Quinn, \emph{{Constraints Imposed by CP Conservation in
  the Presence of Instantons}},
  \href{https://doi.org/10.1103/PhysRevD.16.1791}{\emph{Phys. Rev. D}
  {\bfseries 16} (1977) 1791}.

\bibitem{Weinberg:1977ma}
S.~Weinberg, \emph{{A New Light Boson?}},
  \href{https://doi.org/10.1103/PhysRevLett.40.223}{\emph{Phys. Rev. Lett.}
  {\bfseries 40} (1978) 223}.

\bibitem{Wilczek:1977pj}
F.~Wilczek, \emph{{Problem of Strong $P$ and $T$ Invariance in the Presence of
  Instantons}}, \href{https://doi.org/10.1103/PhysRevLett.40.279}{\emph{Phys.
  Rev. Lett.} {\bfseries 40} (1978) 279}.

\bibitem{DiLuzio:2020wdo}
L.~Di~Luzio, M.~Giannotti, E.~Nardi and L.~Visinelli, \emph{{The landscape of
  QCD axion models}},
  \href{https://doi.org/10.1016/j.physrep.2020.06.002}{\emph{Phys. Rept.}
  {\bfseries 870} (2020) 1} [\href{https://arxiv.org/abs/2003.01100}{{\ttfamily
  2003.01100}}].

\bibitem{Baluni:1978rf}
V.~Baluni, \emph{{CP Violating Effects in QCD}},
  \href{https://doi.org/10.1103/PhysRevD.19.2227}{\emph{Phys. Rev. D}
  {\bfseries 19} (1979) 2227}.

\bibitem{Crewther:1979pi}
R.J.~Crewther, P.~Di~Vecchia, G.~Veneziano and E.~Witten, \emph{{Chiral
  Estimate of the Electric Dipole Moment of the Neutron in Quantum
  Chromodynamics}},
  \href{https://doi.org/10.1016/0370-2693(79)90128-X}{\emph{Phys. Lett. B}
  {\bfseries 88} (1979) 123}.

\bibitem{Kim:1979if}
J.E.~Kim, \emph{{Weak Interaction Singlet and Strong CP Invariance}},
  \href{https://doi.org/10.1103/PhysRevLett.43.103}{\emph{Phys. Rev. Lett.}
  {\bfseries 43} (1979) 103}.

\bibitem{Shifman:1979if}
M.A.~Shifman, A.I.~Vainshtein and V.I.~Zakharov, \emph{{Can Confinement Ensure
  Natural CP Invariance of Strong Interactions?}},
  \href{https://doi.org/10.1016/0550-3213(80)90209-6}{\emph{Nucl. Phys. B}
  {\bfseries 166} (1980) 493}.

\bibitem{Dine:1981rt}
M.~Dine, W.~Fischler and M.~Srednicki, \emph{{A Simple Solution to the Strong
  CP Problem with a Harmless Axion}},
  \href{https://doi.org/10.1016/0370-2693(81)90590-6}{\emph{Phys. Lett. B}
  {\bfseries 104} (1981) 199}.

\bibitem{Zhitnitsky:1980tq}
A.R.~Zhitnitsky, \emph{{On Possible Suppression of the Axion Hadron
  Interactions. (In Russian)}}, {\emph{Sov. J. Nucl. Phys.} {\bfseries 31}
  (1980) 260}.

\bibitem{Baker:2006ts}
C.A.~Baker et~al., \emph{{An Improved experimental limit on the electric dipole
  moment of the neutron}},
  \href{https://doi.org/10.1103/PhysRevLett.97.131801}{\emph{Phys. Rev. Lett.}
  {\bfseries 97} (2006) 131801}
  [\href{https://arxiv.org/abs/hep-ex/0602020}{{\ttfamily hep-ex/0602020}}].

\bibitem{Pendlebury:2015lrz}
J.M.~Pendlebury et~al., \emph{{Revised experimental upper limit on the electric
  dipole moment of the neutron}},
  \href{https://doi.org/10.1103/PhysRevD.92.092003}{\emph{Phys. Rev. D}
  {\bfseries 92} (2015) 092003}
  [\href{https://arxiv.org/abs/1509.04411}{{\ttfamily 1509.04411}}].

\bibitem{Kim:1984pt}
J.E.~Kim, \emph{{A COMPOSITE INVISIBLE AXION}},
  \href{https://doi.org/10.1103/PhysRevD.31.1733}{\emph{Phys. Rev. D}
  {\bfseries 31} (1985) 1733}.

\bibitem{Choi:1985cb}
K.~Choi and J.E.~Kim, \emph{{DYNAMICAL AXION}},
  \href{https://doi.org/10.1103/PhysRevD.32.1828}{\emph{Phys. Rev. D}
  {\bfseries 32} (1985) 1828}.

\bibitem{Kaplan:1985dv}
D.B.~Kaplan, \emph{{Opening the Axion Window}},
  \href{https://doi.org/10.1016/0550-3213(85)90319-0}{\emph{Nucl. Phys. B}
  {\bfseries 260} (1985) 215}.

\bibitem{Randall:1992ut}
L.~Randall, \emph{{Composite axion models and Planck scale physics}},
  \href{https://doi.org/10.1016/0370-2693(92)91928-3}{\emph{Phys. Lett. B}
  {\bfseries 284} (1992) 77}.

\bibitem{Dobrescu:1996jp}
B.A.~Dobrescu, \emph{{The Strong CP problem versus Planck scale physics}},
  \href{https://doi.org/10.1103/PhysRevD.55.5826}{\emph{Phys. Rev. D}
  {\bfseries 55} (1997) 5826}
  [\href{https://arxiv.org/abs/hep-ph/9609221}{{\ttfamily hep-ph/9609221}}].

\bibitem{Redi:2016esr}
M.~Redi and R.~Sato, \emph{{Composite Accidental Axions}},
  \href{https://doi.org/10.1007/JHEP05(2016)104}{\emph{JHEP} {\bfseries 05}
  (2016) 104} [\href{https://arxiv.org/abs/1602.05427}{{\ttfamily
  1602.05427}}].

\bibitem{Kobakhidze:2016wmv}
A.~Kobakhidze, \emph{{Solving the Strong CP Problem with High-Colour Quarks and
  Composite Axion}},  \href{https://arxiv.org/abs/1602.06363}{{\ttfamily
  1602.06363}}.

\bibitem{Dvali:2016eay}
G.~Dvali and L.~Funcke, \emph{{Domestic Axion}},
  \href{https://arxiv.org/abs/1608.08969}{{\ttfamily 1608.08969}}.

\bibitem{Fukuda:2017ylt}
H.~Fukuda, M.~Ibe, M.~Suzuki and T.T.~Yanagida, \emph{{A ''gauged'' $U(1)$
  Peccei\textendash{}Quinn symmetry}},
  \href{https://doi.org/10.1016/j.physletb.2017.05.071}{\emph{Phys. Lett. B}
  {\bfseries 771} (2017) 327}
  [\href{https://arxiv.org/abs/1703.01112}{{\ttfamily 1703.01112}}].

\bibitem{DiLuzio:2017tjx}
L.~Di~Luzio, E.~Nardi and L.~Ubaldi, \emph{{Accidental Peccei-Quinn symmetry
  protected to arbitrary order}},
  \href{https://doi.org/10.1103/PhysRevLett.119.011801}{\emph{Phys. Rev. Lett.}
  {\bfseries 119} (2017) 011801}
  [\href{https://arxiv.org/abs/1704.01122}{{\ttfamily 1704.01122}}].

\bibitem{Lillard:2017cwx}
B.~Lillard and T.M.P.~Tait, \emph{{A Composite Axion from a Supersymmetric
  Product Group}}, \href{https://doi.org/10.1007/JHEP11(2017)005}{\emph{JHEP}
  {\bfseries 11} (2017) 005}
  [\href{https://arxiv.org/abs/1707.04261}{{\ttfamily 1707.04261}}].

\bibitem{DiLuzio:2017pfr}
L.~Di~Luzio, F.~Mescia and E.~Nardi, \emph{{Window for preferred axion
  models}}, \href{https://doi.org/10.1103/PhysRevD.96.075003}{\emph{Phys. Rev.
  D} {\bfseries 96} (2017) 075003}
  [\href{https://arxiv.org/abs/1705.05370}{{\ttfamily 1705.05370}}].

\bibitem{Gaillard:2018xgk}
M.K.~Gaillard, M.B.~Gavela, R.~Houtz, P.~Quilez and R.~Del~Rey, \emph{{Color
  unified dynamical axion}},
  \href{https://doi.org/10.1140/epjc/s10052-018-6396-6}{\emph{Eur. Phys. J. C}
  {\bfseries 78} (2018) 972}
  [\href{https://arxiv.org/abs/1805.06465}{{\ttfamily 1805.06465}}].

\bibitem{Lillard:2018fdt}
B.~Lillard and T.M.P.~Tait, \emph{{A High Quality Composite Axion}},
  \href{https://doi.org/10.1007/JHEP11(2018)199}{\emph{JHEP} {\bfseries 11}
  (2018) 199} [\href{https://arxiv.org/abs/1811.03089}{{\ttfamily
  1811.03089}}].

\bibitem{Anastasopoulos:2018uyu}
P.~Anastasopoulos, P.~Betzios, M.~Bianchi, D.~Consoli and E.~Kiritsis,
  \emph{{Emergent/Composite axions}},
  \href{https://doi.org/10.1007/JHEP10(2019)113}{\emph{JHEP} {\bfseries 10}
  (2019) 113} [\href{https://arxiv.org/abs/1811.05940}{{\ttfamily
  1811.05940}}].

\bibitem{Gavela:2018paw}
M.B.~Gavela, M.~Ibe, P.~Quilez and T.T.~Yanagida, \emph{{Automatic
  Peccei\textendash{}Quinn symmetry}},
  \href{https://doi.org/10.1140/epjc/s10052-019-7046-3}{\emph{Eur. Phys. J. C}
  {\bfseries 79} (2019) 542}
  [\href{https://arxiv.org/abs/1812.08174}{{\ttfamily 1812.08174}}].

\bibitem{Lee:2018yak}
H.-S.~Lee and W.~Yin, \emph{{Peccei-Quinn symmetry from a hidden gauge group
  structure}}, \href{https://doi.org/10.1103/PhysRevD.99.015041}{\emph{Phys.
  Rev. D} {\bfseries 99} (2019) 015041}
  [\href{https://arxiv.org/abs/1811.04039}{{\ttfamily 1811.04039}}].

\bibitem{Dienes:1999gw}
K.R.~Dienes, E.~Dudas and T.~Gherghetta, \emph{{Invisible axions and large
  radius compactifications}},
  \href{https://doi.org/10.1103/PhysRevD.62.105023}{\emph{Phys. Rev. D}
  {\bfseries 62} (2000) 105023}
  [\href{https://arxiv.org/abs/hep-ph/9912455}{{\ttfamily hep-ph/9912455}}].

\bibitem{Hill:2002kq}
C.T.~Hill and A.K.~Leibovich, \emph{{Natural Theories of Ultralow Mass PNGB's:
  Axions and Quintessence}},
  \href{https://doi.org/10.1103/PhysRevD.66.075010}{\emph{Phys. Rev. D}
  {\bfseries 66} (2002) 075010}
  [\href{https://arxiv.org/abs/hep-ph/0205237}{{\ttfamily hep-ph/0205237}}].

\bibitem{Choi:2003wr}
K.-w.~Choi, \emph{{A QCD axion from higher dimensional gauge field}},
  \href{https://doi.org/10.1103/PhysRevLett.92.101602}{\emph{Phys. Rev. Lett.}
  {\bfseries 92} (2004) 101602}
  [\href{https://arxiv.org/abs/hep-ph/0308024}{{\ttfamily hep-ph/0308024}}].

\bibitem{Flacke:2006ad}
T.~Flacke, B.~Gripaios, J.~March-Russell and D.~Maybury, \emph{{Warped
  axions}}, \href{https://doi.org/10.1088/1126-6708/2007/01/061}{\emph{JHEP}
  {\bfseries 01} (2007) 061}
  [\href{https://arxiv.org/abs/hep-ph/0611278}{{\ttfamily hep-ph/0611278}}].

\bibitem{Cox:2019rro}
P.~Cox, T.~Gherghetta and M.D.~Nguyen, \emph{{A Holographic Perspective on the
  Axion Quality Problem}},
  \href{https://doi.org/10.1007/JHEP01(2020)188}{\emph{JHEP} {\bfseries 01}
  (2020) 188} [\href{https://arxiv.org/abs/1911.09385}{{\ttfamily
  1911.09385}}].

\bibitem{Bonnefoy:2020llz}
Q.~Bonnefoy, P.~Cox, E.~Dudas, T.~Gherghetta and M.D.~Nguyen, \emph{{Flavoured
  Warped Axion}}, \href{https://doi.org/10.1007/JHEP04(2021)084}{\emph{JHEP}
  {\bfseries 04} (2021) 084}
  [\href{https://arxiv.org/abs/2012.09728}{{\ttfamily 2012.09728}}].

\bibitem{Yamada:2021uze}
M.~Yamada and T.T.~Yanagida, \emph{{A natural and simple UV completion of the
  QCD axion model}},
  \href{https://doi.org/10.1016/j.physletb.2021.136267}{\emph{Phys. Lett. B}
  {\bfseries 816} (2021) 136267}
  [\href{https://arxiv.org/abs/2101.10350}{{\ttfamily 2101.10350}}].

\bibitem{Wise:1981ry}
M.B.~Wise, H.~Georgi and S.L.~Glashow, \emph{{SU(5) and the Invisible Axion}},
  \href{https://doi.org/10.1103/PhysRevLett.47.402}{\emph{Phys. Rev. Lett.}
  {\bfseries 47} (1981) 402}.

\bibitem{Lazarides:1981kz}
G.~Lazarides, \emph{{SO(10) and the Invisible Axion}},
  \href{https://doi.org/10.1103/PhysRevD.25.2425}{\emph{Phys. Rev. D}
  {\bfseries 25} (1982) 2425}.

\bibitem{Georgi:1981pu}
H.M.~Georgi, L.J.~Hall and M.B.~Wise, \emph{{Grand Unified Models With an
  Automatic {Peccei-Quinn} Symmetry}},
  \href{https://doi.org/10.1016/0550-3213(81)90433-8}{\emph{Nucl. Phys. B}
  {\bfseries 192} (1981) 409}.

\bibitem{Rubakov:1997vp}
V.A.~Rubakov, \emph{{Grand unification and heavy axion}},
  \href{https://doi.org/10.1134/1.567390}{\emph{JETP Lett.} {\bfseries 65}
  (1997) 621} [\href{https://arxiv.org/abs/hep-ph/9703409}{{\ttfamily
  hep-ph/9703409}}].

\bibitem{Co:2016xti}
R.T.~Co, F.~D'Eramo and L.J.~Hall, \emph{{Supersymmetric axion grand unified
  theories and their predictions}},
  \href{https://doi.org/10.1103/PhysRevD.94.075001}{\emph{Phys. Rev. D}
  {\bfseries 94} (2016) 075001}
  [\href{https://arxiv.org/abs/1603.04439}{{\ttfamily 1603.04439}}].

\bibitem{Lee:2016wiy}
C.-H.~Lee and R.N.~Mohapatra, \emph{{Vector-Like Quarks and Leptons, SU(5)
  $\otimes$ SU(5) Grand Unification, and Proton Decay}},
  \href{https://doi.org/10.1007/JHEP02(2017)080}{\emph{JHEP} {\bfseries 02}
  (2017) 080} [\href{https://arxiv.org/abs/1611.05478}{{\ttfamily
  1611.05478}}].

\bibitem{Boucenna:2017fna}
S.M.~Boucenna and Q.~Shafi, \emph{{Axion inflation, proton decay, and
  leptogenesis in $SU(5)\times U(1)_{PQ}$}},
  \href{https://doi.org/10.1103/PhysRevD.97.075012}{\emph{Phys. Rev. D}
  {\bfseries 97} (2018) 075012}
  [\href{https://arxiv.org/abs/1712.06526}{{\ttfamily 1712.06526}}].

\bibitem{Daido:2018dmu}
R.~Daido, F.~Takahashi and N.~Yokozaki, \emph{{Enhanced
  axion\textendash{}photon coupling in GUT with hidden photon}},
  \href{https://doi.org/10.1016/j.physletb.2018.03.039}{\emph{Phys. Lett. B}
  {\bfseries 780} (2018) 538}
  [\href{https://arxiv.org/abs/1801.10344}{{\ttfamily 1801.10344}}].

\bibitem{DiLuzio:2018gqe}
L.~Di~Luzio, A.~Ringwald and C.~Tamarit, \emph{{Axion mass prediction from
  minimal grand unification}},
  \href{https://doi.org/10.1103/PhysRevD.98.095011}{\emph{Phys. Rev. D}
  {\bfseries 98} (2018) 095011}
  [\href{https://arxiv.org/abs/1807.09769}{{\ttfamily 1807.09769}}].

\bibitem{Ernst:2018rod}
A.~Ernst, L.~Di~Luzio, A.~Ringwald and C.~Tamarit, \emph{{Axion properties in
  GUTs}}, \href{https://doi.org/10.22323/1.347.0054}{\emph{PoS} {\bfseries
  CORFU2018} (2019) 054} [\href{https://arxiv.org/abs/1811.11860}{{\ttfamily
  1811.11860}}].

\bibitem{FileviezPerez:2019fku}
P.~Fileviez~P\'erez, C.~Murgui and A.D.~Plascencia, \emph{{The QCD Axion and
  Unification}}, \href{https://doi.org/10.1007/JHEP11(2019)093}{\emph{JHEP}
  {\bfseries 11} (2019) 093}
  [\href{https://arxiv.org/abs/1908.01772}{{\ttfamily 1908.01772}}].

\bibitem{FileviezPerez:2019ssf}
P.~Fileviez~P\'erez, C.~Murgui and A.D.~Plascencia, \emph{{Axion Dark Matter,
  Proton Decay and Unification}},
  \href{https://doi.org/10.1007/JHEP01(2020)091}{\emph{JHEP} {\bfseries 01}
  (2020) 091} [\href{https://arxiv.org/abs/1911.05738}{{\ttfamily
  1911.05738}}].

\bibitem{Bajc:2005zf}
B.~Bajc, A.~Melfo, G.~Senjanovic and F.~Vissani, \emph{{Yukawa sector in
  non-supersymmetric renormalizable SO(10)}},
  \href{https://doi.org/10.1103/PhysRevD.73.055001}{\emph{Phys. Rev. D}
  {\bfseries 73} (2006) 055001}
  [\href{https://arxiv.org/abs/hep-ph/0510139}{{\ttfamily hep-ph/0510139}}].

\bibitem{Altarelli:2013aqa}
G.~Altarelli and D.~Meloni, \emph{{A non supersymmetric SO(10) grand unified
  model for all the physics below $M_{GUT}$}},
  \href{https://doi.org/10.1007/JHEP08(2013)021}{\emph{JHEP} {\bfseries 08}
  (2013) 021} [\href{https://arxiv.org/abs/1305.1001}{{\ttfamily 1305.1001}}].

\bibitem{Babu:2015bna}
K.S.~Babu and S.~Khan, \emph{{Minimal nonsupersymmetric $SO(10)$ model: Gauge
  coupling unification, proton decay, and fermion masses}},
  \href{https://doi.org/10.1103/PhysRevD.92.075018}{\emph{Phys. Rev. D}
  {\bfseries 92} (2015) 075018}
  [\href{https://arxiv.org/abs/1507.06712}{{\ttfamily 1507.06712}}].

\bibitem{Ernst:2018bib}
A.~Ernst, A.~Ringwald and C.~Tamarit, \emph{{Axion Predictions in $SO(10)\times
  U(1)_{\rm PQ}$ Models}},
  \href{https://doi.org/10.1007/JHEP02(2018)103}{\emph{JHEP} {\bfseries 02}
  (2018) 103} [\href{https://arxiv.org/abs/1801.04906}{{\ttfamily
  1801.04906}}].

\bibitem{Coriano:2019vjl}
C.~Corian\`o, P.H.~Frampton, A.~Tatullo and D.~Theofilopoulos, \emph{{An
  axion-like particle from an $SO(10)$ seesaw with $U (1)_X$}},
  \href{https://doi.org/10.1016/j.physletb.2020.135273}{\emph{Phys. Lett. B}
  {\bfseries 802} (2020) 135273}
  [\href{https://arxiv.org/abs/1906.05810}{{\ttfamily 1906.05810}}].

\bibitem{DiLuzio:2020qio}
L.~Di~Luzio, \emph{{Accidental SO(10) axion from gauged flavour}},
  \href{https://doi.org/10.1007/JHEP11(2020)074}{\emph{JHEP} {\bfseries 11}
  (2020) 074} [\href{https://arxiv.org/abs/2008.09119}{{\ttfamily
  2008.09119}}].

\bibitem{Coriano:2017ghp}
C.~Coriano and P.H.~Frampton, \emph{{Dark Matter as Ultralight Axion-Like
  particle in $E_6 \times U(1)_X$ GUT with QCD Axion}},
  \href{https://doi.org/10.1016/j.physletb.2018.05.067}{\emph{Phys. Lett. B}
  {\bfseries 782} (2018) 380}
  [\href{https://arxiv.org/abs/1712.03865}{{\ttfamily 1712.03865}}].

\bibitem{Chen:2021haa}
N.~Chen, Y.~Liu and Z.~Teng, \emph{{An axion model with the ${\rm SU}(6)$
  unification}},  \href{https://arxiv.org/abs/2106.00223}{{\ttfamily
  2106.00223}}.

\bibitem{Witten:1984dg}
E.~Witten, \emph{{Some Properties of O(32) Superstrings}},
  \href{https://doi.org/10.1016/0370-2693(84)90422-2}{\emph{Phys. Lett. B}
  {\bfseries 149} (1984) 351}.

\bibitem{Kallosh:1995hi}
R.~Kallosh, A.D.~Linde, D.A.~Linde and L.~Susskind, \emph{{Gravity and global
  symmetries}}, \href{https://doi.org/10.1103/PhysRevD.52.912}{\emph{Phys. Rev.
  D} {\bfseries 52} (1995) 912}
  [\href{https://arxiv.org/abs/hep-th/9502069}{{\ttfamily hep-th/9502069}}].

\bibitem{Svrcek:2006yi}
P.~Svrcek and E.~Witten, \emph{{Axions In String Theory}},
  \href{https://doi.org/10.1088/1126-6708/2006/06/051}{\emph{JHEP} {\bfseries
  06} (2006) 051} [\href{https://arxiv.org/abs/hep-th/0605206}{{\ttfamily
  hep-th/0605206}}].

\bibitem{Dimopoulos:1979pp}
S.~Dimopoulos, \emph{{A Solution of the Strong {CP} Problem in Models With
  Scalars}}, \href{https://doi.org/10.1016/0370-2693(79)91233-4}{\emph{Phys.
  Lett. B} {\bfseries 84} (1979) 435}.

\bibitem{Tye:1981zy}
S.H.H.~Tye, \emph{{A Superstrong Force With a Heavy Axion}},
  \href{https://doi.org/10.1103/PhysRevLett.47.1035}{\emph{Phys. Rev. Lett.}
  {\bfseries 47} (1981) 1035}.

\bibitem{Holdom:1982ex}
B.~Holdom and M.E.~Peskin, \emph{{Raising the Axion Mass}},
  \href{https://doi.org/10.1016/0550-3213(82)90228-0}{\emph{Nucl. Phys. B}
  {\bfseries 208} (1982) 397}.

\bibitem{Srednicki:1985xd}
M.~Srednicki, \emph{{Axion Couplings to Matter. 1. CP Conserving Parts}},
  \href{https://doi.org/10.1016/0550-3213(85)90054-9}{\emph{Nucl. Phys. B}
  {\bfseries 260} (1985) 689}.

\bibitem{Flynn:1987rs}
J.M.~Flynn and L.~Randall, \emph{{A Computation of the Small Instanton
  Contribution to the Axion Potential}},
  \href{https://doi.org/10.1016/0550-3213(87)90089-7}{\emph{Nucl. Phys. B}
  {\bfseries 293} (1987) 731}.

\bibitem{Kamionkowski:1992mf}
M.~Kamionkowski and J.~March-Russell, \emph{{Planck scale physics and the
  Peccei-Quinn mechanism}},
  \href{https://doi.org/10.1016/0370-2693(92)90492-M}{\emph{Phys. Lett. B}
  {\bfseries 282} (1992) 137}
  [\href{https://arxiv.org/abs/hep-th/9202003}{{\ttfamily hep-th/9202003}}].

\bibitem{Berezhiani:2000gh}
Z.~Berezhiani, L.~Gianfagna and M.~Giannotti, \emph{{Strong CP problem and
  mirror world: The Weinberg-Wilczek axion revisited}},
  \href{https://doi.org/10.1016/S0370-2693(00)01392-7}{\emph{Phys. Lett. B}
  {\bfseries 500} (2001) 286}
  [\href{https://arxiv.org/abs/hep-ph/0009290}{{\ttfamily hep-ph/0009290}}].

\bibitem{Hsu:2004mf}
S.D.H.~Hsu and F.~Sannino, \emph{{New solutions to the strong CP problem}},
  \href{https://doi.org/10.1016/j.physletb.2004.11.040}{\emph{Phys. Lett. B}
  {\bfseries 605} (2005) 369}
  [\href{https://arxiv.org/abs/hep-ph/0408319}{{\ttfamily hep-ph/0408319}}].

\bibitem{Hook:2014cda}
A.~Hook, \emph{{Anomalous solutions to the strong CP problem}},
  \href{https://doi.org/10.1103/PhysRevLett.114.141801}{\emph{Phys. Rev. Lett.}
  {\bfseries 114} (2015) 141801}
  [\href{https://arxiv.org/abs/1411.3325}{{\ttfamily 1411.3325}}].

\bibitem{Alonso-Alvarez:2018irt}
G.~Alonso-\'Alvarez, M.B.~Gavela and P.~Quilez, \emph{{Axion couplings to
  electroweak gauge bosons}},
  \href{https://doi.org/10.1140/epjc/s10052-019-6732-5}{\emph{Eur. Phys. J. C}
  {\bfseries 79} (2019) 223}
  [\href{https://arxiv.org/abs/1811.05466}{{\ttfamily 1811.05466}}].

\bibitem{Hook:2019qoh}
A.~Hook, S.~Kumar, Z.~Liu and R.~Sundrum, \emph{{High Quality QCD Axion and the
  LHC}}, \href{https://doi.org/10.1103/PhysRevLett.124.221801}{\emph{Phys. Rev.
  Lett.} {\bfseries 124} (2020) 221801}
  [\href{https://arxiv.org/abs/1911.12364}{{\ttfamily 1911.12364}}].

\bibitem{Turner:1989vc}
M.S.~Turner, \emph{{Windows on the Axion}},
  \href{https://doi.org/10.1016/0370-1573(90)90172-X}{\emph{Phys. Rept.}
  {\bfseries 197} (1990) 67}.

\bibitem{Fukuda:2015ana}
H.~Fukuda, K.~Harigaya, M.~Ibe and T.T.~Yanagida, \emph{{Model of visible QCD
  axion}}, \href{https://doi.org/10.1103/PhysRevD.92.015021}{\emph{Phys. Rev.
  D} {\bfseries 92} (2015) 015021}
  [\href{https://arxiv.org/abs/1504.06084}{{\ttfamily 1504.06084}}].

\bibitem{Gherghetta:2016fhp}
T.~Gherghetta, N.~Nagata and M.~Shifman, \emph{{A Visible QCD Axion from an
  Enlarged Color Group}},
  \href{https://doi.org/10.1103/PhysRevD.93.115010}{\emph{Phys. Rev. D}
  {\bfseries 93} (2016) 115010}
  [\href{https://arxiv.org/abs/1604.01127}{{\ttfamily 1604.01127}}].

\bibitem{Dimopoulos:2016lvn}
S.~Dimopoulos, A.~Hook, J.~Huang and G.~Marques-Tavares, \emph{{A collider
  observable QCD axion}},
  \href{https://doi.org/10.1007/JHEP11(2016)052}{\emph{JHEP} {\bfseries 11}
  (2016) 052} [\href{https://arxiv.org/abs/1606.03097}{{\ttfamily
  1606.03097}}].

\bibitem{Chiang:2016eav}
C.-W.~Chiang, H.~Fukuda, M.~Ibe and T.T.~Yanagida, \emph{{750 GeV diphoton
  resonance in a visible heavy QCD axion model}},
  \href{https://doi.org/10.1103/PhysRevD.93.095016}{\emph{Phys. Rev. D}
  {\bfseries 93} (2016) 095016}
  [\href{https://arxiv.org/abs/1602.07909}{{\ttfamily 1602.07909}}].

\bibitem{Gherghetta:2020ofz}
T.~Gherghetta and M.D.~Nguyen, \emph{{A Composite Higgs with a Heavy Composite
  Axion}}, \href{https://doi.org/10.1007/JHEP12(2020)094}{\emph{JHEP}
  {\bfseries 12} (2020) 094}
  [\href{https://arxiv.org/abs/2007.10875}{{\ttfamily 2007.10875}}].

\bibitem{Kobakhidze:2016rwh}
A.~Kobakhidze, \emph{{Heavy axion in asymptotically safe QCD}},
  \href{https://arxiv.org/abs/1607.06552}{{\ttfamily 1607.06552}}.

\bibitem{Agrawal:2017ksf}
P.~Agrawal and K.~Howe, \emph{{Factoring the Strong CP Problem}},
  \href{https://doi.org/10.1007/JHEP12(2018)029}{\emph{JHEP} {\bfseries 12}
  (2018) 029} [\href{https://arxiv.org/abs/1710.04213}{{\ttfamily
  1710.04213}}].

\bibitem{Gherghetta:2020keg}
T.~Gherghetta, V.V.~Khoze, A.~Pomarol and Y.~Shirman, \emph{{The Axion Mass
  from 5D Small Instantons}},
  \href{https://doi.org/10.1007/JHEP03(2020)063}{\emph{JHEP} {\bfseries 03}
  (2020) 063} [\href{https://arxiv.org/abs/2001.05610}{{\ttfamily
  2001.05610}}].

\bibitem{Choi:2020rgn}
K.~Choi, S.H.~Im and C.S.~Shin, \emph{{Recent progress in physics of axions or
  axion-like particles}},  \href{https://arxiv.org/abs/2012.05029}{{\ttfamily
  2012.05029}}.

\bibitem{Jaeckel:2012yz}
J.~Jaeckel, M.~Jankowiak and M.~Spannowsky, \emph{{LHC probes the hidden
  sector}}, \href{https://doi.org/10.1016/j.dark.2013.06.001}{\emph{Phys. Dark
  Univ.} {\bfseries 2} (2013) 111}
  [\href{https://arxiv.org/abs/1212.3620}{{\ttfamily 1212.3620}}].

\bibitem{Brivio:2017ije}
I.~Brivio, M.B.~Gavela, L.~Merlo, K.~Mimasu, J.M.~No, R.~del Rey et~al.,
  \emph{{ALPs Effective Field Theory and Collider Signatures}},
  \href{https://doi.org/10.1140/epjc/s10052-017-5111-3}{\emph{Eur. Phys. J. C}
  {\bfseries 77} (2017) 572}
  [\href{https://arxiv.org/abs/1701.05379}{{\ttfamily 1701.05379}}].

\bibitem{Bauer:2017ris}
M.~Bauer, M.~Neubert and A.~Thamm, \emph{{Collider Probes of Axion-Like
  Particles}}, \href{https://doi.org/10.1007/JHEP12(2017)044}{\emph{JHEP}
  {\bfseries 12} (2017) 044}
  [\href{https://arxiv.org/abs/1708.00443}{{\ttfamily 1708.00443}}].

\bibitem{Mariotti:2017vtv}
A.~Mariotti, D.~Redigolo, F.~Sala and K.~Tobioka, \emph{{New LHC bound on
  low-mass diphoton resonances}},
  \href{https://doi.org/10.1016/j.physletb.2018.06.039}{\emph{Phys. Lett. B}
  {\bfseries 783} (2018) 13}
  [\href{https://arxiv.org/abs/1710.01743}{{\ttfamily 1710.01743}}].

\bibitem{Bauer:2018uxu}
M.~Bauer, M.~Heiles, M.~Neubert and A.~Thamm, \emph{{Axion-Like Particles at
  Future Colliders}},
  \href{https://doi.org/10.1140/epjc/s10052-019-6587-9}{\emph{Eur. Phys. J. C}
  {\bfseries 79} (2019) 74} [\href{https://arxiv.org/abs/1808.10323}{{\ttfamily
  1808.10323}}].

\bibitem{Gavela:2019cmq}
M.B.~Gavela, J.M.~No, V.~Sanz and J.F.~de~Troc\'oniz, \emph{{Nonresonant
  Searches for Axionlike Particles at the LHC}},
  \href{https://doi.org/10.1103/PhysRevLett.124.051802}{\emph{Phys. Rev. Lett.}
  {\bfseries 124} (2020) 051802}
  [\href{https://arxiv.org/abs/1905.12953}{{\ttfamily 1905.12953}}].

\bibitem{Carmona:2021seb}
A.~Carmona, C.~Scherb and P.~Schwaller, \emph{{Charming ALPs}},
  \href{https://doi.org/10.1007/JHEP08(2021)121}{\emph{JHEP} {\bfseries 08}
  (2021) 121} [\href{https://arxiv.org/abs/2101.07803}{{\ttfamily
  2101.07803}}].

\bibitem{Florez:2021zoo}
A.~Fl\'orez, A.~Gurrola, W.~Johns, P.~Sheldon, E.~Sheridan, K.~Sinha et~al.,
  \emph{{Probing axionlike particles with $\gamma\gamma$ final states from
  vector boson fusion processes at the LHC}},
  \href{https://doi.org/10.1103/PhysRevD.103.095001}{\emph{Phys. Rev. D}
  {\bfseries 103} (2021) 095001}
  [\href{https://arxiv.org/abs/2101.11119}{{\ttfamily 2101.11119}}].

\bibitem{Wang:2021uyb}
D.~Wang, L.~Wu, J.M.~Yang and M.~Zhang, \emph{{Photon-jet as a probe of
  axion-like particles at the LHC}},
  \href{https://arxiv.org/abs/2102.01532}{{\ttfamily 2102.01532}}.

\bibitem{Ren:2021prq}
J.~Ren, D.~Wang, L.~Wu, J.M.~Yang and M.~Zhang, \emph{{Detecting an axion-like
  particle with machine learning at the LHC}},
  \href{https://arxiv.org/abs/2106.07018}{{\ttfamily 2106.07018}}.

\bibitem{Liu:2017zdh}
J.~Liu, L.-T.~Wang, X.-P.~Wang and W.~Xue, \emph{{Exposing the dark sector with
  future Z factories}},
  \href{https://doi.org/10.1103/PhysRevD.97.095044}{\emph{Phys. Rev. D}
  {\bfseries 97} (2018) 095044}
  [\href{https://arxiv.org/abs/1712.07237}{{\ttfamily 1712.07237}}].

\bibitem{Zhang:2021sio}
H.-Y.~Zhang, C.-X.~Yue, Y.-C.~Guo and S.~Yang, \emph{{Searching for axion-like
  particles at future electron-positron colliders}},
  \href{https://arxiv.org/abs/2103.05218}{{\ttfamily 2103.05218}}.

\bibitem{Cacciapaglia:2021aqz}
G.~Cacciapaglia, A.~Deandrea, A.M.~Iyer and K.~Sridhar, \emph{{Tera-Zooming in
  on light (composite) axion-like particles}},
  \href{https://arxiv.org/abs/2104.11064}{{\ttfamily 2104.11064}}.

\bibitem{Steinberg:2021wbs}
N.~Steinberg, \emph{{Discovering Axion-Like Particles with Photon Fusion at the
  ILC}},  \href{https://arxiv.org/abs/2108.11927}{{\ttfamily 2108.11927}}.

\bibitem{dEnterria:2021ljz}
D.~d'Enterria, \emph{{Collider constraints on axion-like particles}},  in
  \emph{{Workshop on Feebly Interacting Particles}}, 2, 2021
  [\href{https://arxiv.org/abs/2102.08971}{{\ttfamily 2102.08971}}].

\bibitem{Apollinari:2015wtw}
G.~Apollinari, O.~Br\"uning, T.~Nakamoto and L.~Rossi, \emph{{High Luminosity
  Large Hadron Collider HL-LHC}},
  \href{https://doi.org/10.5170/CERN-2015-005.1}{\emph{CERN Yellow Rep.} (2015)
  1} [\href{https://arxiv.org/abs/1705.08830}{{\ttfamily 1705.08830}}].

\bibitem{FCC:2018byv}
{\scshape FCC} collaboration, \emph{{FCC Physics Opportunities}: {Future
  Circular Collider Conceptual Design Report Volume 1}},
  \href{https://doi.org/10.1140/epjc/s10052-019-6904-3}{\emph{Eur. Phys. J. C}
  {\bfseries 79} (2019) 474}.

\bibitem{CEPC-SPPCStudyGroup:2015csa}
M.~Ahmad et~al., \emph{{CEPC-SPPC Preliminary Conceptual Design Report. 1.
  Physics and Detector}}, .

\bibitem{muCcoll:2020}
``Muon collider collaboration meeting.''
  \url{https://indico.cern.ch/event/930508/}.

\bibitem{MICE:2019jkl}
{\scshape MICE} collaboration, \emph{{Demonstration of cooling by the Muon
  Ionization Cooling Experiment}},
  \href{https://doi.org/10.1038/s41586-020-1958-9}{\emph{Nature} {\bfseries
  578} (2020) 53} [\href{https://arxiv.org/abs/1907.08562}{{\ttfamily
  1907.08562}}].

\bibitem{Delahaye:2019omf}
J.P.~Delahaye, M.~Diemoz, K.~Long, B.~Mansouli\'e, N.~Pastrone, L.~Rivkin
  et~al., \emph{{Muon Colliders}},
  \href{https://arxiv.org/abs/1901.06150}{{\ttfamily 1901.06150}}.

\bibitem{Bartosik:2020xwr}
N.~Bartosik et~al., \emph{{Detector and Physics Performance at a Muon
  Collider}},
  \href{https://doi.org/10.1088/1748-0221/15/05/P05001}{\emph{JINST} {\bfseries
  15} (2020) P05001} [\href{https://arxiv.org/abs/2001.04431}{{\ttfamily
  2001.04431}}].

\bibitem{Linssen:2012hp}
L.~Linssen, A.~Miyamoto, M.~Stanitzki and H.~Weerts, eds., \emph{{Physics and
  Detectors at CLIC: CLIC Conceptual Design Report}},
  \href{https://arxiv.org/abs/1202.5940}{{\ttfamily 1202.5940}}.

\bibitem{Han:2020uid}
T.~Han, Y.~Ma and K.~Xie, \emph{{High Energy Leptonic Collisions and
  Electroweak Parton Distribution Functions}},
  \href{https://arxiv.org/abs/2007.14300}{{\ttfamily 2007.14300}}.

\bibitem{Kilian:2007gr}
W.~Kilian, T.~Ohl and J.~Reuter, \emph{{WHIZARD: Simulating Multi-Particle
  Processes at LHC and ILC}},
  \href{https://doi.org/10.1140/epjc/s10052-011-1742-y}{\emph{Eur. Phys. J. C}
  {\bfseries 71} (2011) 1742}
  [\href{https://arxiv.org/abs/0708.4233}{{\ttfamily 0708.4233}}].

\bibitem{Alwall:2014hca}
J.~Alwall, R.~Frederix, S.~Frixione, V.~Hirschi, F.~Maltoni, O.~Mattelaer
  et~al., \emph{{The automated computation of tree-level and next-to-leading
  order differential cross sections, and their matching to parton shower
  simulations}}, \href{https://doi.org/10.1007/JHEP07(2014)079}{\emph{JHEP}
  {\bfseries 07} (2014) 079} [\href{https://arxiv.org/abs/1405.0301}{{\ttfamily
  1405.0301}}].

\end{thebibliography}\endgroup

\end{document}